\documentclass{svmult}
\usepackage{amsmath,amsfonts,amssymb,dsfont,graphicx,makeidx,psfrag}
\newtheorem{exception}{Exception}
\newcommand{\ii}{\mathrm{i}}
\newcommand{\ij}{\mathrm{j}}
\DeclareMathOperator{\cs}{cs}
\DeclareMathOperator{\e}{e}

\begin{document}

\title*{Numerical computation of Maass waveforms and an application to cosmology}
\titlerunning{Computation of Maass waveforms and an application to cosmology}
\author{R.~Aurich\inst{1}, F.~Steiner\inst{1}, and H.~Then\inst{2}}
\institute{Abteilung~Theoretische~Physik, Universit\"{a}t~Ulm, Albert-Einstein-Allee~11, 89069~Ulm, Germany, \texttt{ralf.aurich@physik.uni-ulm.de}, \texttt{frank.steiner@physik.uni-ulm.de} \and School~of~Mathematics, University~of~Bristol, University~Walk, Bristol, BS8~1TW, United~Kingdom, \texttt{holger.then@bristol.ac.uk}}

\date{\small February 2004}

\maketitle

\begin{abstract} We compute numerically eigenvalues and eigenfunctions of the Laplacian in a three-dimensional hyperbolic space. Applying the results to cosmology, we demonstrate that the methods learned in quantum chaos can be used in other fields of research. \end{abstract}

\section{Introduction} \label{sec:1}
\begin{figure}
\centering
\input{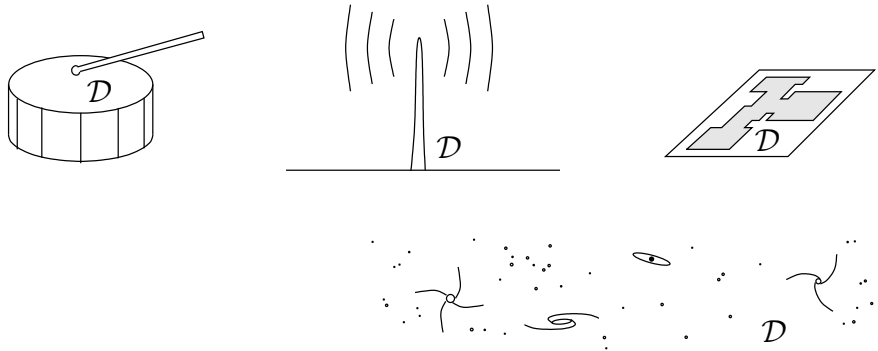}
\caption{Applications of the Laplacian. A drum (top left), electromagnetic waves (top middle), semiconductors resp. quantum mechanics (top right), and the universe (bottom).}
\label{fig:1.1}
\end{figure}
The Laplacian $\Delta$, a second order differential operator, is of fundamental interest in several fields of mathematics and physics. One of the oldest examples is a drum ${\cal D}$, see figure \ref{fig:1.1} top left. If one hits it, its membrane oscillates and gives a sound. The vibrations of the drum are solutions of the Helmholtz equation $(c=1)$,
\begin{align} (\Delta-\frac{\partial^2}{\partial t^2})u(x,t)=0 \quad \forall x\in{\cal D}, \label{eq:1.1} \end{align}
where $u$ is the displacement of the membrane. Fixing the membrane to its frame gives rise to Dirichlet boundary conditions,
\begin{align} u(x,t)=0 \quad \forall x\in\partial{\cal D}. \label{eq:1.2} \end{align}
A Fourier transformation,
\begin{align} u(x,t)=\int_{-\infty}^{\infty} v(x,\omega) \e^{-\ii\omega t} \, d\omega, \label{eq:1.3} \end{align}
allows us to eliminate the time-dependence. The sound of the drum is determined by the eigenvalues of the time-independent Helmholtz equation,
\begin{align} (\Delta+\omega^2)v(x,\omega)=0 \quad \forall x\in{\cal D}, \label{eq:1.4} \end{align}
which is nothing else than the eigenvalue equation of the (negative) Laplacian inside the domain ${\cal D}$. \par
Another example is the propagation of electromagnetic waves in a two-dimensional system ${\cal D}$, see figure \ref{fig:1.1} top middle. Each component of the electric and magnetic field is again subject to the Helmholtz equation (\ref{eq:1.1}). Placing (super)conducting materials around ${\cal D}$ yields Dirichlet boundary conditions (\ref{eq:1.2}). Electromagnetic waves inside ${\cal D}$ can only be transmitted and received, if their frequencies are in the spectrum of the Laplacian. \par
Of special interest to this proceedings is quantum chaos which yields our next example, see figure \ref{fig:1.1} top right. A non-relativistic point particle moving freely in a manifold resp. orbifold ${\cal D}$ is described by the Schr\"{o}dinger equation
\begin{align} \ii\hbar\frac{\partial}{\partial t}\Psi(x,t)=-\frac{\hbar^2}{2m}\Delta\Psi(x,t) \quad \forall x\in{\cal D}, \label{eq:1.5} \end{align}
with appropriate boundary conditions on $\partial{\cal D}$. Scaling the units to $\hbar=2m=1$, and making the ansatz
\begin{align} \Psi(x,t)=\psi(x)\e^{-\frac{\ii}{\hbar}Et}, \label{eq:1.8} \end{align}
gives the time-independent Schr\"{o}dinger equation,
\begin{align} (\Delta+E)\psi(x)=0 \quad \forall x\in{\cal D}. \label{eq:1.9} \end{align}
The statistical properties of its spectrum and its eigenfunctions are a central subject of study in quantum chaos. \par
In sections \ref{sec:8} and \ref{sec:8a} we present some of the statistical properties of the solutions of the eigenvalue equation (\ref{eq:1.9}). But before, in sections \ref{sec:3}--\ref{sec:5} we introduce the three-dimensional hyperbolic system we are dealing with, and in sections \ref{sec:6} and \ref{sec:7} we develop an efficient algorithm that allows us to compute the solutions numerically. \par
From section \ref{sec:9} on we apply the results of the eigenvalue equation of the Laplacian to the universe, see figure \ref{fig:1.1} bottom, and compute the temperature fluctuations in the cosmic microwave background (CMB). This final example demonstrates that the methods learned in quantum chaos can be successfully used in other fields of research even though the physical interpretation can differ completely, e.g.\ metric perturbations and temperature fluctuations in cosmology instead of probability amplitudes in quantum mechanics. \par

\section{General statistical properties in quantum chaos} \label{sec:2} Concerning the statistical properties of the eigenvalues and the eigenfunctions, we have to emphasise that they depend on the choice of the manifold resp. orbifold ${\cal D}$. Depending on whether the corresponding classical system is integrable or not, there are some generally accepted conjectures about the nearest-neighbour spacing distributions of the eigenvalues in the semiclassical limit. The semiclassical limit is the limit of large eigenvalues, $E\to\infty$. \par
Unless otherwise stated, we use the following assumptions:
The quantum mechanical system is desymmetrised with respect to all
its unitary symmetries, and whenever we examine the distribution of
the eigenvalues we regard them on the scale of the mean level spacings.
Moreover, it is generally believed that after desymmetrisation a
generic quantum Hamiltonian corresponding to a classically strongly chaotic 
system possesses no degenerate eigenvalues.
\begin{conjecture}[Berry, Tabor \cite{BerryTabor1976}] \label{conj:2.1}
If the corresponding classical system is integrable, the eigenvalues
behave like independent random variables and the distribution of the
nearest-neighbour spacings is in the semiclassical limit close to the
Poisson distribution, i.e.\ there is no level repulsion.
\end{conjecture}
\begin{conjecture}[Bohigas, Giannoni, Schmit \cite{BohigasGiannoniSchmit1984,
BohigasGiannoniSchmit1986}] \label{conj:2.2}
If the corresponding classical system is chaotic, the eigenvalues are
distributed like the eigenvalues of hermitian random matrices
\cite{Dyson1970,Mehta1991}. The corresponding ensembles depend only on the
symmetries of the system:
\begin{itemize}
\item
For chaotic systems without time-reversal invariance the distribution
of the eigenvalues approaches in the semiclassical limit the distribution
of the Gauss\-ian Unitary Ensemble (GUE) which is characterised by a
quadratic level repulsion.
\item
For chaotic systems with time-reversal invariance and integer spin
the distribution of the eigenvalues approaches in the semiclassical limit
the distribution of the Gauss\-ian Orthogonal Ensemble (GOE) which is
characterised by a linear level repulsion.
\item
For chaotic systems with time-reversal invariance and half-integer
spin the distribution of the eigenvalues approaches in the semiclassical
limit the distribution of the Gauss\-ian Symplectic Ensemble (GSE) which is
characterised by a quartic level repulsion.
\end{itemize}
\end{conjecture}
These conjectures are very well confirmed by numerical calculations,
but several exceptions are known. Here are two examples:
\begin{exception} \label{exept:2.1}
The harmonic oscillator is classically integrable, but its spectrum
is equidistant.
\end{exception}
\begin{exception} \label{exept:2.2}
The geodesic motion on surfaces with constant negative curvature
provides a prime example for classical chaos. In some cases, however,
the nearest-neighbour distribution of the eigenvalues of the Laplacian
on these surfaces appears to be Poissonian.
\end{exception}
``A strange arithmetical structure of chaos'' in the case of surfaces of
constant negative curvature that are generated by arithmetic fundamental
groups was discovered by Aurich and Steiner \cite{AurichSteiner1988}, see
also Aurich, Bogomolny, and Steiner \cite{AurichBogomolnySteiner1991}.
(For the definition of an arithmetic group we refer the reader to
\cite{Borel1969}). Deviations from the expected GOE-behaviour
in the case of a particular arithmetic surface were numerically
observed by Bohigas, Giannoni, and Schmit
\cite{BohigasGiannoniSchmit1986} and by Aurich and Steiner
\cite{AurichSteiner1989}. Computations showed
\cite{AurichSteiner1989,AurichSteiner1990}, however, that the
level statistics on $30$ generic (i.e.\ non-arithmetic) surfaces were in
nice agreement with the expected random-matrix theory prediction in
accordance with conjecture \ref{conj:2.2}. This has led Bogomolny,
Georgeot, Giannoni, and Schmit \cite{BogomolnyGeorgeotGiannoniSchmit1992},
Bolte, Steil, and Steiner \cite{BolteSteilSteiner1992}, and Sarnak
\cite{Sarnak1995} to introduce the concept of arithmetic quantum chaos.
\begin{conjecture}[Arithmetic Quantum Chaos] \label{conj:2.3}
On surfaces of constant negative curvature that are generated by
arithmetic fundamental groups, the distribution of the eigenvalues of the
quantum Hamiltonian approaches in the semiclassical limit the Poisson
distribution. Due to level clustering small spacings occur comparably often.
\end{conjecture} \par
In order to carry out some specific numerical computations, we have to specify the manifold resp. orbifold ${\cal D}$. We will choose ${\cal D}$ to be given by the quotient space $\Gamma\backslash{\cal H}$ in the hyperbolic upper half space ${\cal H}$ where we choose $\Gamma$ to be the Picard group, see section \ref{sec:5}.

\section{The hyperbolic upper half-space} \label{sec:3} 
\begin{figure}
\centering
\includegraphics[width=5cm,height=8cm,angle=-90]{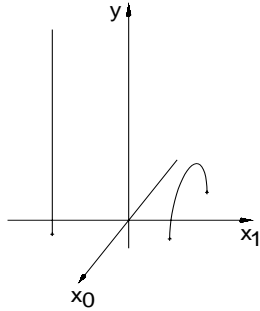}
\caption{Geodesics in the upper half-space of constant negative
curvature.}
\label{fig:3.1}
\end{figure}

Let
\begin{align} {\cal H}=\{(x_0,x_1,y)\in\mathds{R}^3; \quad y>0 \} \label{eq:3.1} \end{align}
be the upper half-space equipped with the hyperbolic metric of constant 
curvature $-1$
\begin{align}
ds^2=\frac{dx_0^2+dx_1^2+dy^2}{y^2}. \label{eq:3.2}
\end{align}
Due to the metric the Laplacian reads
\begin{align} \Delta=y^2\big(\frac{\partial^2}{\partial x_0^2}+\frac{\partial^2}{\partial x_1^2}+\frac{\partial^2}{\partial y^2}\big)-y\frac{\partial}{\partial y}, \label{eq:3.3} \end{align}
and the volume element is
\begin{align} d\mu=\frac{dx_0 dx_1 dy}{y^3}. \label{eq:3.4} \end{align}
The geodesics of a particle moving freely in the upper half-space
are straight lines and semicircles perpendicular to the
$x_0$-$x_1$-plane, respectively, see figure \ref{fig:3.1}.
 \par
Expressing any point $(x_0,x_1,y)\in{\cal H}$ as a Hamilton
quaternion, $z=x_0+\ii x_1+\ij y$, with the multiplication defined
by $\ii^2=-1,\ \ij^2=-1,\ \ii\ij+\ij\ii=0$, all motions in the upper
half-space are given by linear fractional transformations
\begin{align}
z\mapsto\gamma z=(az+b)(cz+d)^{-1}; \quad a,b,c,d\in\mathds{C}, \quad
ad-bc=1. \label{eq:3.5}
\end{align}
The group of these transformations is isomorphic to the group of matrices
\begin{align}
\gamma=\begin{pmatrix}a&b\\c&d\end{pmatrix}\in
\operatorname{SL}(2,\mathds{C}) \label{eq:3.6}
\end{align}
up to a common sign of the matrix entries,
\begin{align}
\operatorname{SL}(2,\mathds{C})/\{\pm1\}
=\operatorname{PSL}(2,\mathds{C}). \label{eq:3.7}
\end{align}
The motions provided by the elements of $\operatorname{PSL}(2,\mathds{C})$
exhaust all orientation preserving isometries of the hyperbolic metric
on ${\cal H}$.
\begin{remark} \label{rem:3.1}
If one wants to avoid using quaternions, the point
$(x_0,x_1,y)\in{\cal H}$ can be expressed by
$(x,y)\in\mathds{C}\times\mathds{R}$ with $x=x_0+\ii x_1$ and $y>0$.
But then the linear fractional transformations look somewhat more
complicated,
\begin{align}
(x,y)\mapsto\gamma(x,y)=\big(
\frac{(ax+b)(\bar{c}\bar{x}+\bar{d})+a\bar{c}y^2}{|cx+d|^2+|cy|^2},
\frac{y}{|cx+d|^2+|cy|^2}\big). \label{eq:3.8}
\end{align}
In order to keep the notation simple, we mainly use quaternions.
\end{remark}

\section{Topology} \label{sec:4} The topology is given by the manifold resp. orbifold. An orbifold is a space that locally looks like a Euclidean space modulo the action of a discrete group, e.g.\ a rotation group. A manifold is an orbifold that locally resembles Euclidean space. \par
A hyperbolic three-orbifold can be realised by a quotient
\begin{align} \Gamma\backslash{\cal H}=\{\Gamma z; \quad z\in{\cal H}\}, \label{eq:4.1} \end{align}
where ${\cal H}$ is the upper-half space and $\Gamma$ is a discrete subgroup of the isometries on ${\cal H}$. The elements of $\Gamma$ are $2\times2$-matrices whose determinants equal one. If the trace of an element is real, it is called hyperbolic, parabolic, or elliptic, depending on whether the absolute value of its trace is larger, equal, or smaller than two, respectively. If the trace of an element is not real, the element is called loxodromic. \par
The action of $\Gamma$ on ${\cal H}$ identifies all $\Gamma$-equivalent points with each other. All the $\Gamma$-equivalent points of $z$ give an orbit
\begin{align} \Gamma z=\{\gamma z; \quad \gamma\in\Gamma\}. \label{eq:4.3} \end{align}
The set of all the orbits is the orbifold. If, except of the identity, $\Gamma$ contains only parabolic and hyperbolic elements, then $\Gamma\backslash{\cal H}$ is a manifold. \par
There exist infinitely many hyperbolic three-orbifolds. But opposed to the two-dimensional case there do not exist any deformations of hyperbolic three-orbifolds, because of the Mostow rigidity theorem \cite{Mostow1973,Prasad1973}. \par
Each hyperbolic three-orbifold has its specific volume which can be finite or infinite. The volumes are always bounded from below by a positive constant. \par
If the volume of $\Gamma\backslash{\cal H}$ is finite and if $\Gamma$ does not contain any parabolic elements, then the orbifold is compact. If $\Gamma$ does contain parabolic elements, then the orbifold has cusps and is non-compact. \par
The hyperbolic three-manifold of smallest volume is unknown. Only a lower limit for the volume is proven to be \cite{Przeworski2001}
\begin{align} \operatorname{vol}({\cal M})>0.281. \label{eq:4.4} \end{align}
The smallest known hyperbolic three-manifold is the Weeks manifold \cite{Weeks1985} whose volume is
\begin{align} \operatorname{vol}({\cal M})\simeq0.943. \label{eq:4.5} \end{align}
The volume of the Weeks manifold is the smallest among the volumes of arithmetic hyperbolic three-manifolds \cite{ChinburgFriedmanJonesReid2001}. \par
In contrast, hyperbolic three-orbifolds are known which have smaller volumes, whereby the one with smallest volume is also unknown. A lower limit for the volume is \cite{Meyerhoff1988a,Meyerhoff1988b}
\begin{align} \operatorname{vol}(\Gamma\backslash{\cal H})>8.2\cdot10^{-4}, \label{eq:4.6} \end{align}
and the smallest known hyperbolic orbifold is the twofold extension of the tetrahedral Coxeter group $\operatorname{CT}(22)$. With
\begin{align} \Gamma=\operatorname{CT}(22)_2^+=\operatorname{CT}(22)_2\cap\operatorname{Iso}^+({\cal H}), \label{eq:4.7} \end{align}
where $\operatorname{Iso}^+({\cal H})$ are the orientation preserving isometries of ${\cal H}$, the volume of the orbifold is
\begin{align} \operatorname{vol}(\Gamma\backslash{\cal H})\simeq0.039. \label{eq:4.8} \end{align}
The volume of this orbifold is the smallest among the volumes of arithmetic hyperbolic three-orbifolds \cite{ChinburgFriedman1986}. \par

\section{The Picard group} \label{sec:5} In the following we choose the hyperbolic three-orbifold $\Gamma\backslash{\cal H}$ of constant negative curvature that is generated by the Picard group,
\begin{align} \Gamma=\operatorname{PSL}(2,\mathds{Z}[\ii]), \label{eq:5.1} \end{align}
where $\mathds{Z}[\ii]=\mathds{Z}+\ii\mathds{Z}$ are the Gauss\-ian integers. \par
\begin{figure}
\centering
\includegraphics[width=5cm,height=8cm,angle=-90]{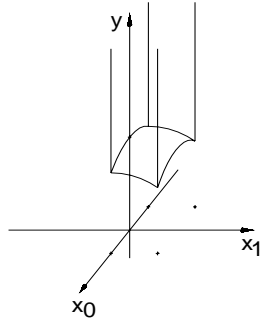}
\caption{The fundamental domain of the Picard group.}
\label{fig:5.1}
\end{figure}
\begin{figure}
\centering
\includegraphics[width=5cm,height=8cm,angle=-90]{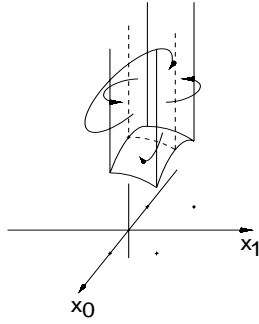}
\caption{Identifying the faces of the fundamental domain according
to the elements of the Picard group.}
\label{fig:5.2}
\end{figure}

The Picard group is generated by the cosets of three elements,
\begin{align} \begin{pmatrix}1&1\\0&1\end{pmatrix}, \quad \begin{pmatrix}1&\ii\\0&1\end{pmatrix}, \quad \begin{pmatrix}0&-1\\1&0\end{pmatrix}, \label{eq:5.2} \end{align}
which yield two translations and one inversion,
\begin{align} z\mapsto z+1, \quad z\mapsto z+\ii, \quad z\mapsto-z^{-1}. \label{eq:5.3} \end{align}
The three motions generating $\Gamma$, together with the coset of the element
\begin{align} \begin{pmatrix}\ii&0\\0&-\ii\end{pmatrix} \label{eq:5.4} \end{align}
that is isomorphic to the symmetry
\begin{align} z=x+\ij y\mapsto\ii z\ii=-x+\ij y, \label{eq:5.5} \end{align}
can be used to construct the fundamental domain.
\begin{definition} \label{def:5.1}
A fundamental domain of the discrete group $\Gamma$ is a closed subset ${\cal F}\subset{\cal H}$ with the following conditions: \\
(i) ${\cal F}$ meets each orbit $\Gamma z$ at least once, \\
(ii) if an orbit $\Gamma z$ does not meet the boundary of ${\cal F}$ it meets ${\cal F}$ at most once, \\
(iii) the boundary of ${\cal F}$ has Lebesgue measure zero. \end{definition}
For the Picard group the fundamental domain of standard shape is
\begin{align} {\cal F}=\{z=x_0+\ii x_1+\ij y\in{\cal H}; \quad -\frac{1}{2}\le x_0\le\frac{1}{2}, \quad 0\le x_1\le\frac{1}{2}, \quad |z|\ge1\}, \label{eq:5.6} \end{align}
with the absolute value of $z$ being defined by $|z|=(x_0^2+x_1^2+y^2)^{\frac{1}{2}}$, see figure \ref{fig:5.1}.
Identifying the faces of the fundamental domain according to the
elements of the group $\Gamma$ leads to a realisation of the
quotient space $\Gamma\backslash{\cal H}$, see figure \ref{fig:5.2}.
\par
With the hyperbolic metric the quotient space $\Gamma\backslash{\cal H}$
inherits the structure of an orbifold that has one parabolic and four
elliptic fix-points,
\begin{align}
z=\ij\infty, \quad z=\ij, \quad z=\frac{1}{2}+\ij\sqrt{\frac{3}{4}},
\quad z=\frac{1}{2}+\ii\frac{1}{2}+\ij\sqrt{\frac{1}{2}}, \quad
z=\ii\frac{1}{2}+\ij\sqrt{\frac{3}{4}}. \label{eq:5.7}
\end{align}
The parabolic fix-point corresponds to a cusp at $z=\ij\infty$ that
is invariant under the parabolic elements
\begin{align}
\begin{pmatrix}1&1\\0&1\end{pmatrix} \quad \text{and} \quad
\begin{pmatrix}1&\ii\\0&1\end{pmatrix}. \label{eq:5.8}
\end{align}
Because of the hyperbolic metric the volume of the non-compact orbifold
$\Gamma\backslash{\cal H}$ is finite \cite{Humbert1919},
\begin{align}
\operatorname{vol}(\Gamma\backslash{\cal H})=\frac{\zeta_K(2)}{4\pi^2}
=0.30532186472\ldots, \label{eq:5.9}
\end{align}
where
\begin{align}
\zeta_K(s)=\frac{1}{4}\sum_{\nu\in\mathds{Z}[\ii]-\{0\}}
(\nu\bar{\nu})^{-s}, \quad \Re s>1, \label{eq:5.10}
\end{align}
is the Dedekind zeta function.

\section{Maass waveforms} \label{sec:6} We are interested in the smooth and square-integrable eigenfunctions of the Laplacian in the orbifold $\Gamma\backslash{\cal H}$. A function being defined on the upper half-space that is invariant under all discrete linear fractional transformations,
\begin{align} \psi(\gamma z)=\psi(z) \quad \forall \gamma\in\Gamma, \label{eq:6.1} \end{align}
is called an automorphic function. Any automorphic function can be identified with a function living on the quotient space $\Gamma\backslash{\cal H}$, and vice versa any function being defined on the quotient space can be identified with an automorphic function living in the upper half-space. The functions we are interested in match the definition of Maass waveforms \cite{Maass1949a,Maass1949b}.
\begin{definition} \label{def:6.1} Let $\Gamma$ be a discrete subgroup of the isometries $\Gamma\subset\operatorname{PSL}(2,\mathds{C})$ containing parabolic elements. A function $\psi(z)$ is called a Maass waveform if it is an automorphic eigenfunction of the Laplacian that is smooth and square-integrable on the fundamental domain,
\begin{align}
& \psi(z)\in C^{\infty}({\cal H}), \label{eq:6.2} \\
& \psi(z)\in L^2(\Gamma\backslash{\cal H}), \label{eq:6.3} \\
& (\Delta+E)\psi(z)=0 \quad \forall z\in{\cal H}, \label{eq:6.4} \\
& \psi(\gamma z)=\psi(z) \quad \forall \gamma\in\Gamma,\ z\in{\cal H}. \label{eq:6.5} \end{align}
\end{definition}
Since the Maass waveforms are automorphic with respect to a discrete subgroup $\Gamma$ that contains parabolic elements, they are periodic in the directions perpendicular to the cusp. This allows to expand them into Fourier series. \par
In case of the Picard group we have
\begin{align}
\psi(z)=u(y)+\sum_{\beta\in\mathds{Z}[\ii]-\{0\}}
a_{\beta}yK_{\ii k}(2\pi|\beta|y)\e^{2\pi\ii\Re\beta x},
\label{eq:6.6}
\end{align}
where $\Re\beta x$ is the real part of the complex scalar product $\beta x$, and
\begin{align}
u(y)=\begin{cases} b_0 y^{1+\ii k}+b_1 y^{1-\ii k}&
\text{if $k\not=0$},\\ b_2 y+b_3 y\ln y& \text{if $k=0$}.
\end{cases} \label{eq:6.7}
\end{align}
$K_{\ii k}(t)$ is the K-Bessel function \cite{Watson1944}
whose order is connected with the eigenvalue $E$ by
\begin{align}
E=k^2+1. \label{eq:6.8}
\end{align}
If a Maass waveform vanishes in the cusp,
\begin{align}
\lim_{z\to\ij\infty}\psi(z)=0, \label{eq:6.9}
\end{align}
it is called a Maass cusp form. \par
According to the Roelcke-Selberg spectral resolution of the Laplacian
\cite{Selberg1956,Roelcke1966,Hejhal1983}, its spectrum contains
both a discrete and a continuous part. The discrete part is spanned by the
constant eigenfunction $\psi_{k_0}$ and a countable number of Maass cusp forms
$\psi_{k_1},\psi_{k_2},\psi_{k_3},\ldots$ which we take to be ordered with
increasing eigenvalues, $0=E_{k_0}<E_{k_1}\le E_{k_2}\le E_{k_3}\le\ldots$.
The continuous part of the spectrum $E\ge1$ is spanned by the
Eisenstein series $E(z,1+\ii k)$ which are known analytically
\cite{Kubota1973,ElstrodtGrunewaldMennicke1985}. The Fourier coefficients
of the function $\Lambda_K(1+\ii k)E(z,1+\ii k)$ are given by
\begin{align}
b_0=\Lambda_K(1+\ii k), \quad b_1=\Lambda_K(1-\ii k), \quad
a_{\beta}=2\sum_{\substack{\lambda,\mu\in\mathds{Z}[\ii] \\
\lambda\mu=\beta}} \big|\frac{\lambda}{\mu}\big|^{\ii k}, \label{eq:6.10}
\end{align}
where $\beta\in\mathds{Z}[\ii]-\{0\}$, and
\begin{align}
\Lambda_K(s)=4\pi^{-s}\Gamma(s)\zeta_K(s) \label{eq:6.11}
\end{align}
has an analytic continuation into the complex plane except for a
pole at $s=1$. \par
Defining
\begin{align} \psi^{\text{Eisen}}_k(z)=\frac{\Lambda_K(1+\ii k)E(z,1+\ii k)}{\sqrt{\pi}|\Lambda_K(1+\ii k)|}, \label{eq:6.12} \end{align}
the Eisenstein series $\psi^{\text{Eisen}}_k(z)$ is real. \par
Normalising the Maass cusp forms according to
\begin{align}
\langle\psi_{k_i},\psi_{k_i}\rangle=1, \label{eq:6.13}
\end{align}
we can expand any square integrable function
$\phi\in L^2(\Gamma\backslash{\cal H})$ in terms of the Maass cusp forms and
the Eisenstein series,
\cite{ElstrodtGrunewaldMennicke1998},
\begin{align}
\phi(z)=\sum_{i\ge0}\langle\psi_{k_i},\phi\rangle\psi_{k_i}(z)+\frac{1}{2\pi\ii}\int_{\Re s=1}\langle E(\cdot,s),\phi\rangle E(z,s)\,ds, \label{eq:6.14}
\end{align}
where
\begin{align} \langle\psi,\phi\rangle=\int_{\Gamma\backslash{\cal H}}\bar{\psi}\phi\,d\mu \label{eq:6.14a} \end{align}
is the Petersson scalar product. \par
The discrete eigenvalues and their associated Maass cusp forms are
not known analytically. Thus, one has to calculate them
numerically. Previous calculations of eigenvalues for the Picard
group can be found in \cite{SmotrovGolovchanskii1991,Huntebrinker1996,
GrunewaldHuntebrinker1996,Steil1999}. By making use of the Hecke operators
\cite{Stark1984,SmotrovGolovchanskii1991,Heitkamp1992,HejhalArno1993}
and the multiplicative relations among the coefficients, Steil
\cite{Steil1999} obtained a non-linear system of equations which
allowed him to compute $2545$ consecutive eigenvalues.
Another way is to extend Hejhal's algorithm \cite{Hejhal1999} to
three dimensions \cite{Then2003}. Improving the procedure of finding
the eigenvalues \cite{Then2002}, we computed $13950$ consecutive
eigenvalues and their corresponding eigenfunctions.

\section{Hejhal's algorithm} \label{sec:7} Hejhal found a linear stable algorithm for computing Maass waveforms together with their eigenvalues which he used for groups acting on the two-dimensional hyperbolic plane \cite{Hejhal1999}, see also \cite{SelanderStrombergsson2002,Avelin2002} for some applications. We extend this algorithm which is based on the Fourier expansion and the automorphy condition and apply it to the Picard group acting on the three-dimensional hyperbolic space. For the Picard group no small eigenvalues $0<E=k^2+1<1$ exist \cite{Stramm1994}. Therefore, $k$ is real and the term $u(y)$ in the Fourier expansion of Maass cusp forms vanishes. Due to the exponential decay of the K-Bessel function for large arguments,
\begin{align} K_{\ii k}(t)\sim\sqrt{\frac{\pi}{2t}}\e^{-t} \quad \text{for $t\to\infty$}, \label{eq:7.1} \end{align}
and the polynomial bound of the coefficients \cite{Maass1949b},
\begin{align}
a_{\beta}=O(|\beta|), \quad |\beta|\to\infty, \label{eq:7.2}
\end{align}
the absolutely convergent Fourier expansion can be truncated,
\begin{align}
\psi(z)=\sum_{\substack{\beta\in\mathds{Z}[\ii]-\{0\}\\|\beta|\le M}} a_{\beta}y K_{\ii k}(2\pi|\beta|y)\e^{2\pi\ii\Re\beta x}+[[\varepsilon]], \label{eq:7.3} \end{align}
if we bound $y$ from below. Given $\varepsilon>0$, $k$, and $y$, we determine the smallest $M=M(\varepsilon,k,y)$ such that the inequalities
\begin{align} 2\pi My\ge k \quad \text{and} \quad K_{\ii k}(2\pi My)\le\varepsilon\max_{t}(K_{\ii k}(t)) \label{eq:7.4} \end{align}
hold. Larger $y$ allow smaller $M$. In all remainder terms,
\begin{align} [[\varepsilon]]=\sum_{\substack{\beta\in\mathds{Z}[\ii]-\{0\}\\|\beta|>M}} a_{\beta}yK_{\ii k}(2\pi|\beta|y)\e^{2\pi\ii\Re\beta x}, \label{eq:7.5} \end{align}
the K-Bessel function decays exponentially in $|\beta|$, and already the K-Bessel function of the first summand of the remainder terms is smaller than $\varepsilon$ times most of the K-Bessel functions in the sum of (\ref{eq:7.3}). Thus, the error $[[\varepsilon]]$ does at most marginally exceed $\varepsilon$. The reason why $[[\varepsilon]]$ can exceed $\varepsilon$ somewhat is due to the possibility that the summands in (\ref{eq:7.3}) cancel each other, or that the coefficients in the remainder terms are larger than in (\ref{eq:7.3}). By a finite two-dimensional Fourier transformation the Fourier expansion (\ref{eq:7.3}) is solved for its coefficients
\begin{align} a_{\gamma}yK_{\ii k}(2\pi|\gamma|y)=\frac{1}{(2Q)^2}\sum_{x\in\mathds{X}[\ii]} \psi(x+\ij y)\e^{-2\pi\ii\Re\gamma x}+[[\varepsilon]], \label{eq:7.6} \end{align}
where $\gamma\in\mathds{Z}[\ii]-\{0\}$, and $\mathds{X}[\ii]$ is a two-dimensional equally distributed set of $(2Q)^2$ numbers,
\begin{align} \mathds{X}[\ii]=\{\frac{l_0+\ii l_1}{2Q}; \quad l_j=-Q+\tfrac{1}{2},-Q+\tfrac{3}{2},\ldots,Q-\tfrac{3}{2},Q-\tfrac{1}{2}, \quad j=0,1\}, \label{eq:7.7} \end{align}
with $2Q>M+|\gamma|$. \par
By automorphy we have
\begin{align} \psi(z)=\psi(z^*), \label{eq:7.8} \end{align}
where $z^*$ is the $\Gamma$-pullback of the point $z$ into the fundamental domain ${\cal F}$,
\begin{align} z^*=\gamma z, \quad \gamma\in\Gamma, \quad z^*\in{\cal F}. \label{eq:7.8a} \end{align}
Thus, a Maass cusp form can be approximated by
\begin{align} \psi(x+\ij y)=\psi(x^*+\ij y^*)=\sum_{\substack{\beta\in\mathds{Z}[\ii]-\{0\}\\|\beta|\le M_0}} a_{\beta}y^*K_{\ii k}(2\pi|\beta|y^*)\e^{2\pi\ii\Re\beta x^*}+[[\varepsilon]], \label{eq:7.9} \end{align}
where $y^*$ is always larger or equal than the height $y_0$ of the lowest points of the fundamental domain ${\cal F}$,
\begin{align} y_0=\min_{z\in{\cal F}}(y)=\frac{1}{\sqrt{2}}, \label{eq:7.9a} \end{align}
allowing us to replace $M(\varepsilon,k,y)$ by $M_0=M(\varepsilon,k,y_0)$. \par
Choosing $y$ smaller than $y_0$ the $\Gamma$-pullback $z\mapsto z^*$ of any point into the fundamental domain ${\cal F}$ makes at least once use of the inversion $z\mapsto-z^{-1}$, possibly together with the translations $z\mapsto z+1$ and $z\mapsto z+\ii$. This is called implicit automorphy, since it guarantees the invariance $\psi(z)=\psi(-z^{-1})$. The conditions $\psi(z)=\psi(z+1)$ and $\psi(z)=\psi(z+\ii)$ are automatically satisfied because of the Fourier expansion. \par
Making use of the implicit automorphy by replacing $\psi(x+\ij y)$ in (\ref{eq:7.6}) with the right-hand side of (\ref{eq:7.9}) gives
\begin{multline} a_{\gamma}yK_{\ii k}(2\pi|\gamma|y) \\ =\frac{1}{(2Q)^2}\sum_{x\in\mathds{X}[\ii]}\sum_{\substack{\beta\in\mathds{Z}[\ii]-\{0\}\\|\beta|\le M_0}} a_{\beta}y^*K_{\ii k}(2\pi|\beta|y^*)\e^{2\pi\ii\beta x^*}\e^{-2\pi\ii\Re\gamma x}+[[2\varepsilon]], \label{eq:7.10} \end{multline}
which is the central identity in the algorithm. \par
\begin{figure}
\centering
\includegraphics[width=3.75cm,height=6cm,angle=-90]{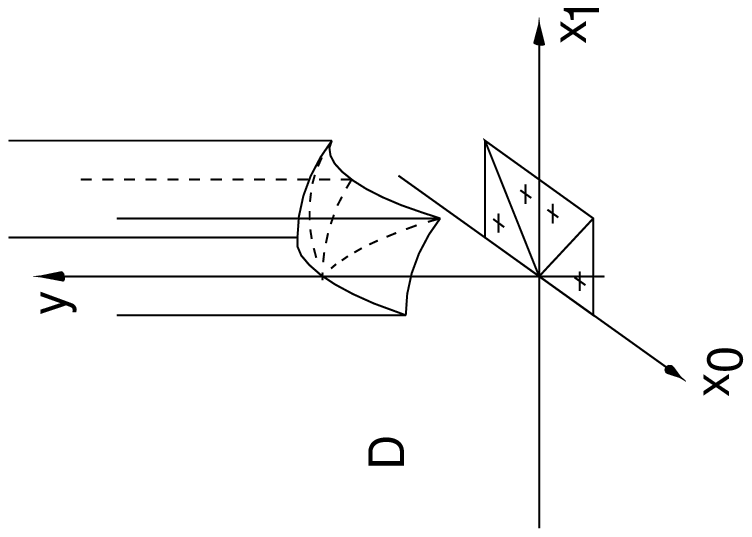} \hspace{-2cm}
\includegraphics[width=3.75cm,height=6cm,angle=-90]{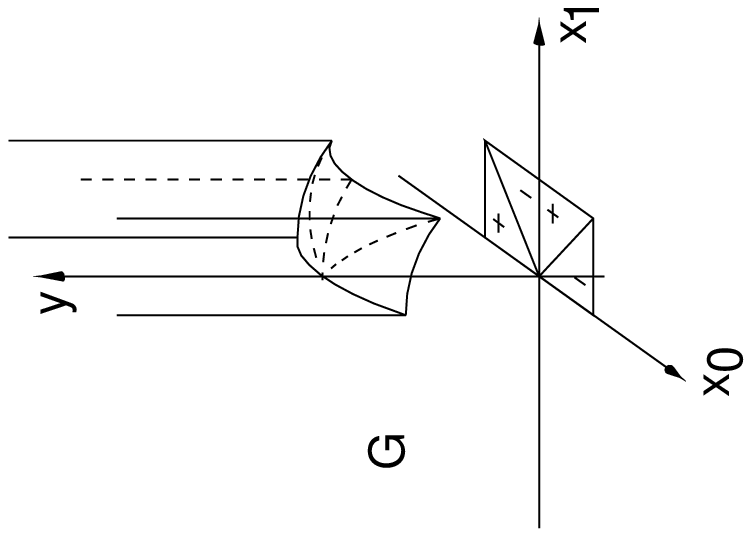} \\
\includegraphics[width=3.75cm,height=6cm,angle=-90]{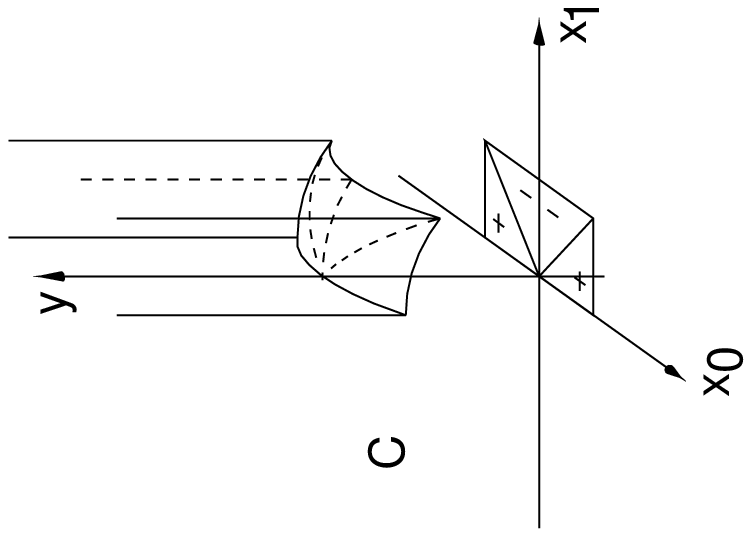} \hspace{-2cm}
\includegraphics[width=3.75cm,height=6cm,angle=-90]{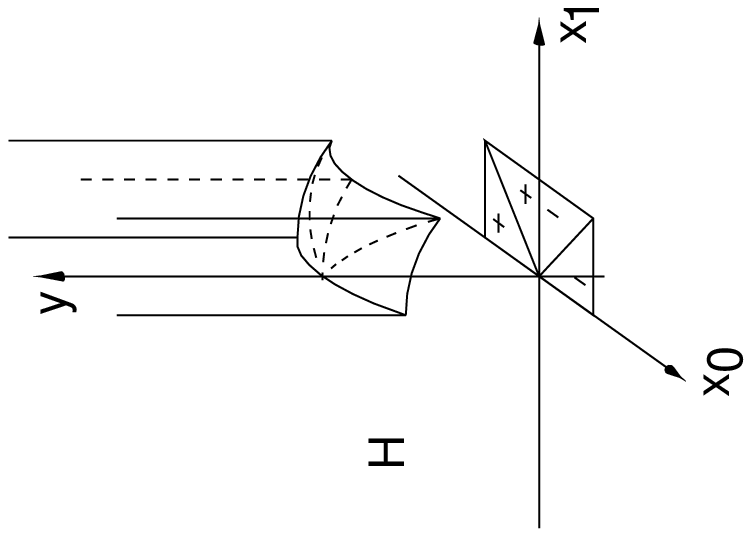}
\caption{The symmetries ${\mathbf D}$, ${\mathbf G}$, ${\mathbf C}$, and ${\mathbf H}$ from top left to bottom right.} \label{fig:7.1} \end{figure}

The symmetry in the Picard group and the symmetries of the fundamental domain imply that the Maass waveforms fall into four symmetry classes \cite{Steil1999} named ${\mathbf D}$, ${\mathbf G}$, ${\mathbf C}$, and ${\mathbf H}$, satisfying
\begin{align} &{\mathbf D}: \quad \psi(x+\ij y)=\psi(\ii x+\ij y)=\psi(-\bar{x}+\ij y),\\ &{\mathbf G}: \quad \psi(x+\ij y)=\psi(\ii x+\ij y)=-\psi(-\bar{x}+\ij y),\\ &{\mathbf C}: \quad \psi(x+\ij y)=-\psi(\ii x+\ij y)=\psi(-\bar{x}+\ij y),\\ &{\mathbf H}: \quad \psi(x+\ij y)=-\psi(\ii x+\ij y)=-\psi(-\bar{x}+\ij y), \label{eq:7.11} \end{align}
respectively, see figure \ref{fig:7.1}, from which the symmetry relations among the coefficients follow,
\begin{align}
&{\mathbf D}: \quad a_{\beta}=a_{\ii\beta}=a_{\bar{\beta}}, \label{eq:7.12} \\
&{\mathbf G}: \quad a_{\beta}=a_{\ii\beta}=-a_{\bar{\beta}}, \label{eq:7.13} \\
&{\mathbf C}: \quad a_{\beta}=-a_{\ii\beta}=a_{\bar{\beta}}, \label{eq:7.14} \\
&{\mathbf H}: \quad a_{\beta}=-a_{\ii\beta}=-a_{\bar{\beta}}. \label{eq:7.15}
\end{align}
Defining
\begin{align} \cs(\beta,x)=\sum_{\sigma\in\mathds{S}_{\beta}} s_{\sigma\beta}\e^{2\pi\ii\Re\sigma x}, \label{eq:7.16} \end{align}
where $s_{\sigma\beta}$ is given by
\begin{align} a_{\sigma}=s_{\sigma\beta}a_{\beta} \label{eq:7.17} \end{align}
and
\begin{align} \sigma\in\mathds{S}_{\beta}=\begin{cases} \{\beta,\ii\beta,-\beta,-\ii\beta,\bar{\beta},\ii\bar{\beta},-\bar{\beta},-\ii\bar{\beta}\}& \text{if $\bar{\beta}\not\in\{\beta,\ii\beta,-\beta,-\ii\beta\}$},\\ \{\beta,\ii\beta,-\beta,-\ii\beta\}& \text{else}, \end{cases} \label{eq:7.18} \end{align}
the Fourier expansion (\ref{eq:6.6}) of the Maass waveforms can be written
\begin{align} \psi(z)=u(y)+\sum_{\beta\in\tilde{\mathds{Z}}[\ii]-\{0\}} a_{\beta}yK_{\ii k}(2\pi|\beta|y)\cs(\beta,x), \label{eq:7.19} \end{align}
where the tilde operator on a set of numbers is defined such that
\begin{align} \tilde{\mathds{X}}\subset\mathds{X}, \quad \bigcup_{x\in\tilde{\mathds{X}}} \mathds{S}_{x}=\mathds{X}, \quad \text{and} \quad \bigcap_{x\in\tilde{\mathds{X}}} \mathds{S}_{x}=\emptyset \label{eq:7.20} \end{align}
holds. \par
Forgetting about the error $[[2\varepsilon]]$ the set of equations (\ref{eq:7.10}) can be written as
\begin{align} \sum_{\substack{\beta\in\tilde{\mathds{Z}}[\ii]-\{0\}\\|\beta|\le M_0}} V_{\gamma\beta}(k,y)a_{\beta}=0, \quad \gamma\in\tilde{\mathds{Z}}[\ii]-\{0\}, \label{eq:7.21} \end{align}
where the matrix $V=(V_{\gamma\beta})$ is given by
\begin{align} V_{\gamma\beta}(k,y)=yK_{\ii k}(2\pi|\gamma|y)\delta_{\gamma\beta}-\frac{1}{(2Q)^2}\sum_{x\in\tilde{\mathds{X}}[\ii]} y^*K_{\ii k}(2\pi|\beta|y^*)\cs(\beta,x^*)\cs(-\gamma,x). \label{eq:7.22} \end{align}
Since $y<y_0$ can always be chosen such that $K_{\ii k}(2\pi|\gamma|y)$ is not too small, the diagonal terms in the matrix $V$ do not vanish for large $|\gamma|$ and the matrix is well conditioned. \par
We are now looking for the non-trivial solutions of (\ref{eq:7.21}) for $1\le|\gamma|\le M_0$ that simultaneously give the eigenvalues $E=k^2+1$ and the coefficients $a_{\beta}$. Trivial solutions are avoided by setting the first non-vanishing coefficient equal to one, $a_{\alpha}=1$, where $\alpha$ is $1$, $2+\ii$, $1$, and $1+\ii$, for the symmetry classes ${\mathbf D}$, ${\mathbf G}$, ${\mathbf C}$, and ${\mathbf H}$, respectively. \par
Since the eigenvalues are unknown, we discretise the $k$-axis and solve for each $k$-value on this grid the inhomogeneous system of equations
\begin{align} \sum_{\substack{\beta\in\tilde{\mathds{Z}}[\ii]-\{0,\alpha\}\\|\beta|\le M_0}} V_{\gamma\beta}(k,y^{\#1})a_{\beta}=-V_{\gamma\alpha}(k,y^{\#1}), \quad 1\le|\gamma|\le M_0, \label{eq:7.23} \end{align}
where $y^{\#1}<y_0$ is chosen such that $K_{\ii k}(2\pi|\gamma|y^{\#1})$ is not too small for $1\le|\gamma|\le M_0$. A good value to try for $y^{\#1}$ is given by $2\pi M_0y^{\#1}=k$. \par
It is important to check whether
\begin{align} g_{\gamma}=\sum_{\substack{\beta\in\tilde{\mathds{Z}}[\ii]-\{0\}\\|\beta|\le M_0}} V_{\gamma\beta}(k,y^{\#2})a_{\beta}, \quad 1\le|\gamma|\le M_0, \label{eq:7.24} \end{align}
vanishes where $y^{\#2}$ is another $y$ value independent of $y^{\#1}$. Only if all $g_{\gamma}$ vanish simultaneously the solution of (\ref{eq:7.23}) is independent of $y$. In this case $E=k^2+1$ is an eigenvalue and the $a_{\beta}$'s are the coefficients of the Fourier expansion of the corresponding Maass cusp form. \par
The probability to find a $k$-value such that all $g_{\gamma}$ vanish simultaneously is zero, because the discrete eigenvalues are of measure zero in the real numbers. Therefore, we make use of the intermediate value theorem where we look for simultaneous sign changes in $g_{\gamma}$ when $k$ is varied. Once we have found them in at least half of the $g_{\gamma}$'s, we have found an interval which contains an eigenvalue with high probability. By some bisection and interpolation we can see if this interval really contains an eigenvalue, and by nesting up the interval until its size tends to zero we obtain the eigenvalue. \par
It is conjectured \cite{BogomolnyGeorgeotGiannoniSchmit1992,BolteSteilSteiner1992,Bolte1993,Sarnak1995} that the eigenvalues of the Laplacian to cusp forms of each particular symmetry class possess a spacing distribution close to that of a Poisson random process, see conjecture \ref{conj:2.3}. One therefore expects that small spacings will occur rather often (due to level clustering). In order not to miss eigenvalues which lie close together, we have to make sure that at least one point of the $k$-grid lies between any two successive eigenvalues. On the other hand, we do not want to waste CPU time if there are large spacings. Therefore, we use an adaptive algorithm which tries to predict the next best $k$-value of the grid. It is based on the observation that the coefficients $a_{\beta}$ of two Maass cusp forms of successive eigenvalues must differ. Assume that two eigenvalues lie close together and that the coefficients of the two Maass cusp forms do not differ much. Numerically then both Maass cusp forms would tend to be similar -- which contradicts the fact that different Maass cusp forms are orthogonal to each other with respect to the Petersson scalar product
\begin{align} \langle\psi_{k_i},\psi_{k_j}\rangle=0, \quad \text{if } k_i\not=k_j. \label{eq:7.25} \end{align}
Maass cusp forms corresponding to different eigenvalues are orthogonal because the Laplacian is an essentially self-adjoint operator. Thus, if successive eigenvalues lie close together, the coefficients $a_{\beta}$ must change fast when varying $k$. In contrast, if successive eigenvalues are separated by large spacings, numerically it turns out that often the coefficients change only slowly upon varying $k$. Defining
\begin{align} \tilde{a}_{\beta}=\frac{a_{\beta}}{\sqrt{\sum_{\substack{\gamma\in\tilde{\mathds{Z}}[\ii]-\{0\} \\ |\gamma|\le M_0}}|a_{\gamma}|^2}}, \quad 1\le|\beta|\le M_0, \label{eq:7.26} \end{align}
our adaptive algorithm predicts the next $k$-value of the grid such that the change in the coefficients is
\begin{align} \sum_{\substack{\beta\in\tilde{\mathds{Z}}[\ii]-\{0\} \\ |\beta|\le M_0}}|\tilde{a}_{\beta}(k_{\text{old}})-\tilde{a}_{\beta}(k_{\text{new}})|^2\approx0.04. \label{eq:7.27} \end{align}
For this prediction, the last step in the $k$-grid together with the last change in the coefficients is used to extrapolate linearly the choice for the next $k$-value of the grid. \par
However the adaptive algorithm is not a rigorous one. Sometimes the prediction of the next $k$-value fails so that it is too close or too far away from the previous one. A small number of small steps does not bother us unless the step size tends to zero. But, if the step size is too large, such that the left-hand side of (\ref{eq:7.27}) exceeds $0.16$, we reduce the step size and try again with a smaller $k$-value. \par
Compared to earlier algorithms, our adaptive one tends to miss significantly less eigenvalues per run.

\section{Eigenvalues} \label{sec:8}
\begin{table} \caption{The first few eigenvalues of the negative Laplacian for the Picard group. Listed is $k$, related to the eigenvalues via $E=k^2+1$.} \label{tab:8.1} \begin{align*} {\mathbf D}& & {\mathbf G}& & {\mathbf C}& & {\mathbf H}& \\ \\
\ 8&.55525104 & & & \ 6&.62211934 \\
11&.10856737 & & & 10&.18079978 \\
12&.86991062 & & & 12&.11527484 & 12&.11527484 \\
14&.07966049 & & & 12&.87936900 \\
15&.34827764 & & & 14&.14833073 \\
15&.89184204 & & & 14&.95244267 & 14&.95244267 \\
17&.33640443 & & & 16&.20759420 \\
17&.45131992 & 17&.45131992 & 16&.99496892 & 16&.99496892 \\
17&.77664065 & & & 17&.86305643 & 17&.86305643 \\
19&.06739052 & & & 18&.24391070 \\
19&.22290266 & & & 18&.83298996 \\
19&.41119126 & & & 19&.43054310 & 19&.43054310 \\
20&.00754583 & & & 20&.30030720 & 20&.30030720 \\
20&.70798880 & 20&.70798880 & 20&.60686743 \\
20&.81526852 & & & 21&.37966055 & 21&.37966055 \\
21&.42887079 & & & 21&.44245892 \\
22&.12230276 & & & 21&.83248972 & 21&.83248972 \\
22&.63055256 & & & 22&.58475297 & 22&.58475297 \\
22&.96230105 & 22&.96230105 & 22&.85429195 \\
23&.49617692 & & & 23&.49768305 & 23&.49768305 \\
23&.52784503 & & & 23&.84275866 \\
23&.88978413 & 23&.88978413 & 23&.89515755 & 23&.89515755 \\
24&.34601664 & & & 24&.42133829 & 24&.42133829 \\
24&.57501426 & & & 25&.03278076 & 25&.03278076 \\
24&.70045917 & & & 25&.42905483 \\
25&.47067539 & & & 25&.77588591 & 25&.77588591 \\
25&.50724616 & & & 26&.03903968 \\
25&.72392169 & 25&.72392169 & 26&.12361823 & 26&.12361823 \\
25&.91864376 & 25&.91864376 & 26&.39170209 \\
26&.42695914 & & & 27&.07065195 & 27&.07065195 \\
27&.03326136 & & & 27&.16341524 & 27&.16341524 \\
27&.14291906 & & & 27&.26799477 & 27&.26799477 \\
27&.14498438 & 27&.14498438 & 27&.89811315 \\
\end{align*} \end{table}
\begin{table} \caption{Some consecutive large eigenvalues of the negative Laplacian for the Picard group. Listed is $k$, related to the eigenvalues via $E=k^2+1$.} \label{tab:8.2} \begin{align*} {\mathbf D}& & {\mathbf G}& & {\mathbf C}& & {\mathbf H}& \\ \\
139&.65419675 & 139&.65419675 & 139&.66399548 & 139&.66399548 \\
139&.65434417 & 139&.65434417 & 139&.66785333 & 139&.66785333 \\
139&.65783548 & 139&.65783548 & 139&.66922266 & 139&.66922266 \\
139&.66104047 & 139&.66104047 & 139&.67870460 & 139&.67870460 \\
139&.67694018 &    &          & 139&.68234200 & 139&.68234200 \\
139&.68162707 & 139&.68162707 & 139&.68424704 & 139&.68424704 \\
139&.68657976 &    &          & 139&.69369972 & 139&.69369972 \\
139&.71803029 & 139&.71803029 & 139&.69413379 & 139&.69413379 \\
139&.72166907 & 139&.72166906 & 139&.69657741 & 139&.69657741 \\
139&.78322452 & 139&.78322452 & 139&.73723373 & 139&.73723373 \\
139&.81928622 & 139&.81928622 & 139&.73828541 & 139&.73828541 \\
139&.81985670 & 139&.81985670 & 139&.74467774 & 139&.74467774 \\
139&.82826034 & 139&.82826034 & 139&.75178180 & 139&.75178180 \\
139&.84250751 &    &          & 139&.75260292 & 139&.75260292 \\
139&.87781072 & 139&.87781072 & 139&.79620628 & 139&.79620628 \\
139&.87805540 &    &          & 139&.80138072 & 139&.80138072 \\
139&.88211647 & 139&.88211647 & 139&.81243991 & 139&.81243991 \\
139&.91782003 & 139&.91782003 & 139&.81312982 & 139&.81312982 \\
139&.91893517 &    &          & 139&.82871870 & 139&.82871870 \\
139&.92397167 & 139&.92397167 & 139&.86401372 & 139&.86401372 \\
139&.92721861 & 139&.92721861 & 139&.86461581 & 139&.86461581 \\
139&.93117207 & 139&.93117207 & 139&.89407865 & 139&.89407865 \\
139&.93149277 & 139&.93149277 & 139&.89914777 & 139&.89914777 \\
139&.94067283 &    &          & 139&.90090849 & 139&.90090849 \\
139&.94396890 & 139&.94396890 & 139&.91635302 & 139&.91635302 \\
139&.95074070 &    &          & 139&.94071729 & 139&.94071729 \\
139&.95124805 & 139&.95124805 & 139&.95080198 & 139&.95080198 \\
139&.99098324 & 139&.99098324 & 139&.97043676 & 139&.97043676 \\
140&.00011792 & 140&.00011792 & 140&.00409202 & 140&.00409202 \\
140&.00109753 & 140&.00109753 & 140&.02733151 & 140&.02733151 \\
140&.00626902 & 140&.00626902 & 140&.04198905 & 140&.04198905 \\
140&.00827516 & 140&.00827516 & 140&.04799273 & 140&.04799273 \\
140&.01679122 & 140&.01679122 & 140&.05764030 & 140&.05764030 \\
\end{align*} \end{table}
We have computed $13950$ consecutive eigenvalues and their corresponding eigenfunctions of the (negative) Laplacian for the Picard group. The smallest non-trivial eigenvalue is $E=k^2+1$ with $k=6.6221193402528$ which is in agreement with the lower bound $E>\frac{2\pi^2}{3}$ \cite{Stramm1994}. Table \ref{tab:8.1} shows the first few eigenvalues of each symmetry class and table \ref{tab:8.2} shows some larger ones. The eigenvalues listed in table \ref{tab:8.1} agree with those of Steil \cite{Steil1999} up to five decimal places. \par
\begin{figure} \centering \includegraphics[width=3.4cm,height=11.7cm,angle=-90]{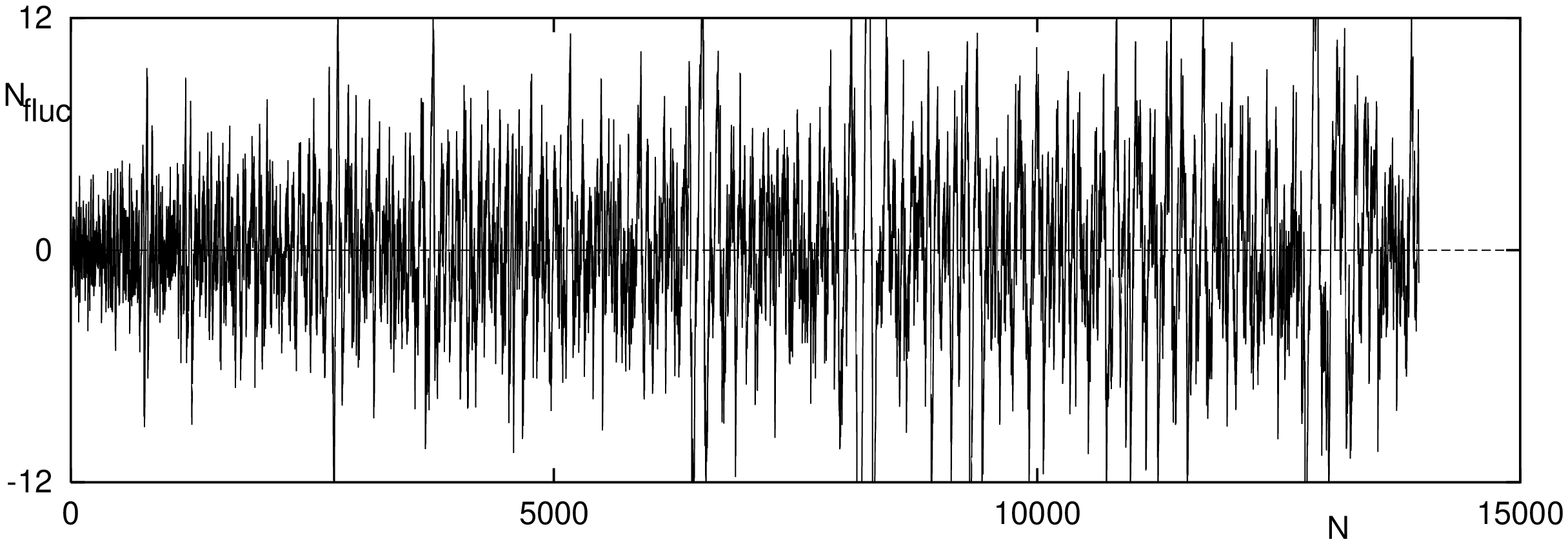} \caption{$N_{fluc}(k_i)$ as a function of $N(k_i)\equiv i$ fluctuating around zero.} \label{fig:8.1} \end{figure}
\begin{figure} \centering \includegraphics[width=3.4cm,height=11.7cm,angle=-90]{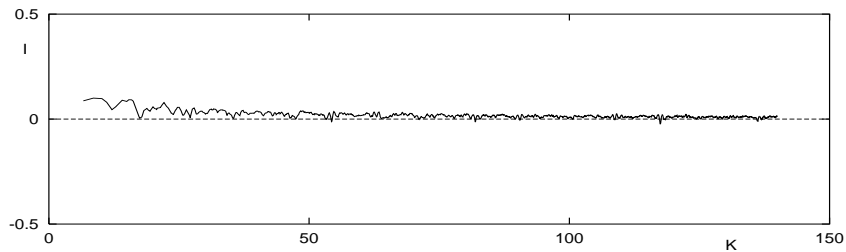} \caption{$I$ as a function of $K$ showing that $I\xrightarrow{K\to\infty}0$.} \label{fig:8.2} \end{figure}

One may ask whether we have found all eigenvalues. The answer can be given by comparing our results with Weyl's law. Consider the level counting function (taking all symmetry classes into account),
\begin{align} N(k)=\#\{\,i\ |\ k_i\le k\}, \label{eq:8.1} \end{align}
(where the trivial eigenvalue, $E=0$, is excluded), and split it into two parts
\begin{align} N(k)=\bar{N}(k)+N_{fluc}(k). \label{eq:8.2} \end{align}
Here $\bar{N}$ is a smooth function describing the average increase in the number of levels, and $N_{fluc}$ describes the fluctuations around the mean such that
\begin{align} \lim_{K\to\infty}\frac{1}{K}\int_{1}^{K}N_{fluc}(k)dk=0. \label{eq:8.3} \end{align}
The average increase in the number of levels is given by Weyl's law \cite{Weyl1912,Avakumovic1956} and higher order corrections have been calculated by Matthies \cite{Matthies1995}. She obtained
\begin{align} \bar{N}(k)=\tfrac{\operatorname{vol}({\cal F})}{6\pi^2}k^3+a_2 k \log k+a_3 k+a_4+o(1) \label{eq:8.4} \end{align}
with the constants
\begin{align} a_2&=-\tfrac{3}{2\pi}, \\ a_3&=\tfrac{1}{\pi}[\tfrac{13}{16}\log 2+\tfrac{7}{4}\log\pi-\log\Gamma(\tfrac{1}{4})+\tfrac{2}{9}\log(2+\sqrt{3})+\tfrac{3}{2}], \\ a_4&=-\tfrac{3}{2}. \label{eq:8.5} \end{align}
We compare our results for $N(k)$ with (\ref{eq:8.4}) by defining
\begin{align} N_{fluc}(k)=N(k)-\bar{N}(k). \label{eq:8.6} \end{align}
$N_{fluc}$ fluctuates around zero or a negative integer whose absolute value gives the number of missing eigenvalues, see figure \ref{fig:8.1}. Unfortunately, our algorithm does not find all eigenvalues in one single run. In the first run it finds about $97\%$ of the eigenvalues. Apart from very few exceptions the remaining eigenvalues are found in the third run. To be more specific, we plotted $N_{fluc}$ decreased by $\frac{1}{2}$, because $N(k)-\bar{N}(k)$ is approximately $\frac{1}{2}$ whenever $E=k^2+1$ is an eigenvalue. A plot indicating that $N_{fluc}$ fluctuates around zero is shown in figure \ref{fig:8.2} where we plotted the integral
\begin{align} I(K)=\frac{1}{K}\int_{1}^{K}N_{fluc}(k)dk. \label{eq:8.7} \end{align}
Desymmetrising the spectrum yields Weyl's law to be
\begin{align} \bar{N}(k)=\frac{\operatorname{vol}({\cal F})}{24\pi^2}k^3+O(k^2) \label{eq:8.8} \end{align}
for each symmetry class. Looking at table \ref{tab:8.1} it seems somehow surprising that there are not equally many eigenvalues listed for each symmetry class. Especially in the symmetry classes ${\mathbf G}$ and ${\mathbf H}$ there seem to be much less eigenvalues than in the symmetry classes ${\mathbf D}$ and ${\mathbf C}$. Indeed, as was shown by Steil \cite{Steil1999}, there occur systematic degenerated eigenvalues between different symmetry classes.
\begin{theorem}[Steil \cite{Steil1999}] \label{thm:8.1} If $E=k^2+1$ is an eigenvalue corresponding to an eigenfunction of the symmetry class ${\mathbf G}$ resp. ${\mathbf H}$, then there exists an eigenfunction of the symmetry class ${\mathbf D}$ resp. ${\mathbf C}$ corresponding to the same eigenvalue. \end{theorem}
Based on our numerical results we conjecture \cite{Then2003}:
\begin{conjecture} \label{conj:8.1} Taking all four symmetry classes together, there are no degenerate eigenvalues other than those explained by Steil's theorem. Furthermore, the degenerate eigenvalues which are explained by Steil's theorem occur only in pairs of two degenerate eigenvalues. They never occur in sets of three or more degenerate eigenvalues. \end{conjecture}
Looking at the semiclassical limit, $E\to\infty$, we finally find that almost all eigenvalues are two-fold degenerated, see e.g.\ table \ref{tab:8.2}, which is an immediate consequence of Weyl's law, Steil's theorem, and conjecture \ref{conj:8.1}. This means that as $E\to\infty$
\begin{align} \frac{\#\{\text{non-degenerate eigenvalues}\le E\}}{\#\{\text{all eigenvalues}\le E\}}\to0. \label{eq:8.9} \end{align} \par
Finally, we remark that the distribution of the eigenvalues of each individual symmetry class agrees numerically with conjecture \ref{conj:2.3}, see \cite{Steil1999,Then2003}.

\section{Eigenfunctions} \label{sec:8a} 
\begin{figure} \psfrag{drho/du}{$\frac{d\rho}{du}$} \psfrag{u}{$u$} \psfrag{rho}{$\rho$} \psfrag{ 0.49999}{} \includegraphics{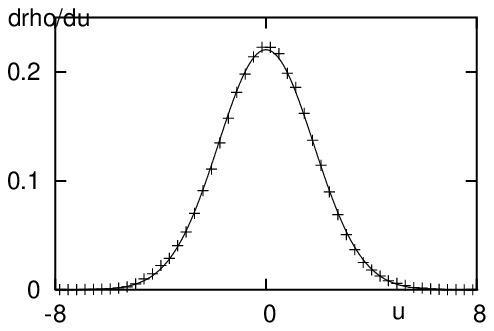} \hspace*{-5cm} \hspace*{\fill} \includegraphics{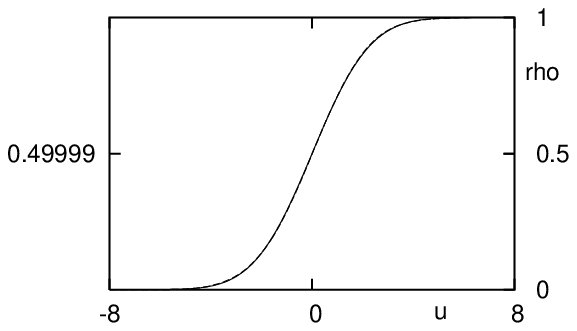} \\ \hspace*{\fill} \includegraphics{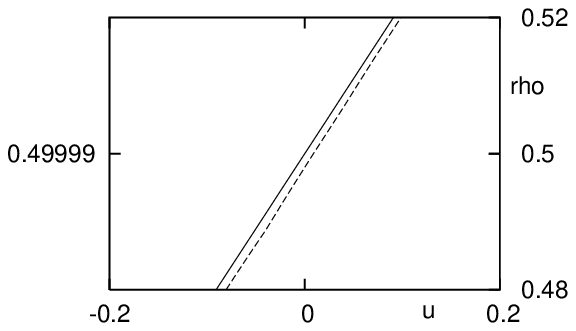} \caption{In the top left figure the value distribution of the eigenfunction corresponding to the eigenvalue $E=k^2+1$ with $k=250.0018575195$ inside the region $F$ is shown as ({\scriptsize$\mbox{}^+$}). The solid line is the conjectured Gauss\-ian. In the figure on the top right the dashed line is the integrated value distribution of the eigenfunction which is nearly indistinguishable from the integrated Gauss\-ian (solid line). The figure on the bottom right displays a detailed magnification of the integrated value distribution showing that it lies slightly below the integrated Gauss\-ian.} \label{fig:8.3} \end{figure}

\begin{figure} \centering \includegraphics[width=11.8cm,angle=0]{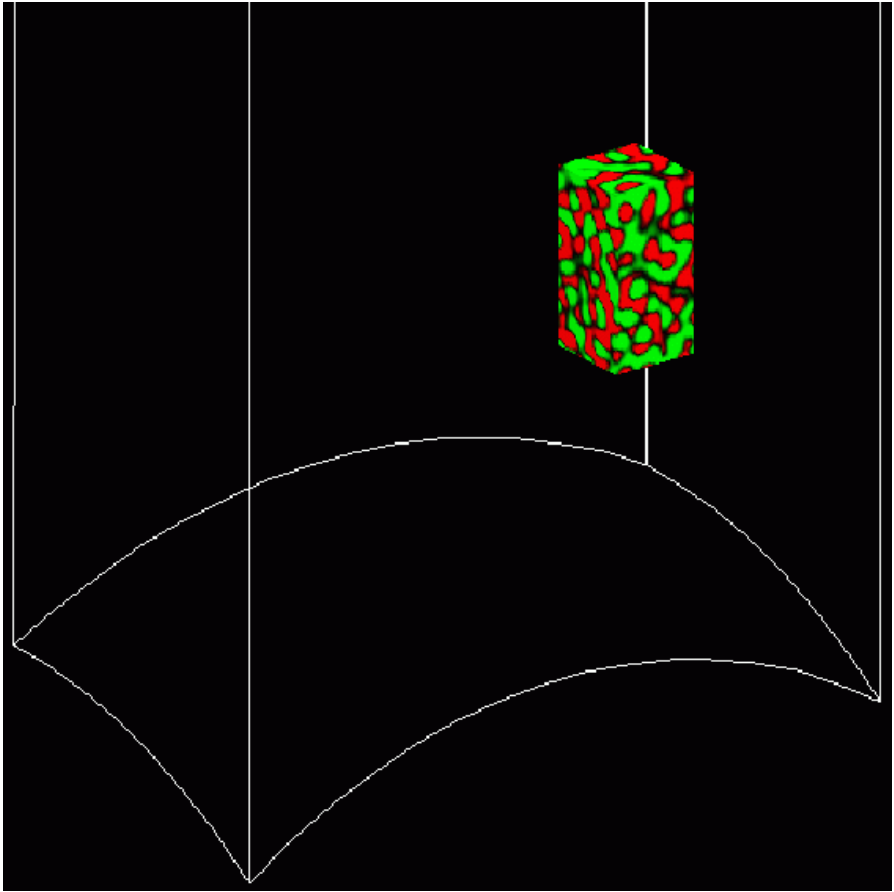} \caption{A plot of the eigenfunction corresponding to the eigenvalue $E=k^2+1$ with $k=250.0018575195$ inside the region $F$.} \label{fig:8.4} \end{figure}

Concerning the eigenfunctions of the Laplacian, it is believed that they behave like random waves. The conjecture of Berry \cite{Berry1977} predicts for each eigenfunction in the semiclassical limit, $E\to\infty$, a Gauss\-ian value distribution,
\begin{align} d\rho(u)=\frac{1}{\sqrt{2\pi}\sigma}\e^{-\frac{u^2}{2\sigma^2}}\,du, \label{eq:8.10} \end{align}
inside any compact regular subregion $F$ of $\cal F$. This means that
\begin{align} \lim_{E\to\infty} \frac{\frac{1}{\operatorname{vol}(F)} \int_F \chi_{[a,b]}(\psi(z))\,d\mu}{\int_a^b\,d\rho(u)}=1 \label{eq:8.11} \end{align}
holds with variance
\begin{align} \sigma^2=\frac{1}{\operatorname{vol}(F)} \int_F |\psi(z)|^2\,d\mu \label{eq:8.12} \end{align}
for any $-\infty<a<b<\infty$, where $\chi_{[a,b]}$ is the indicator function of the interval $[a,b]$. Figure \ref{fig:8.3} shows the value distribution of the $80148$th\footnote{The number $80148$ was determined approximately using Weyl's law (\ref{eq:8.4}).} eigenfunction corresponding to the eigenvalue $E=k^2+1$ with $k=250.0018575195$ inside a small subregion
\begin{align} F=\{z=x+\ii y; \quad -0.43750<x_0<-0.31435, \nonumber \\ 0.06250<x_1<0.18565, \label{eq:8.13} \\ 1.10000<y<1.34881\}. \nonumber \end{align}
Our numerical data agree quite well with Berry's conjecture, providing numerical evidence that the conjecture holds. A plot of the eigenfunction inside the region $F$ is given in figure \ref{fig:8.4}.

\section{An application to cosmology} \label{sec:9} In the remaining sections we apply the eigenvalues and eigenfunctions of the Laplacian to a perturbed Ro\-bert\-son-Wal\-ker universe and compute the temperature fluctuations in the cosmic microwave background (CMB).
\begin{figure} \centering \includegraphics[height=15.2cm]{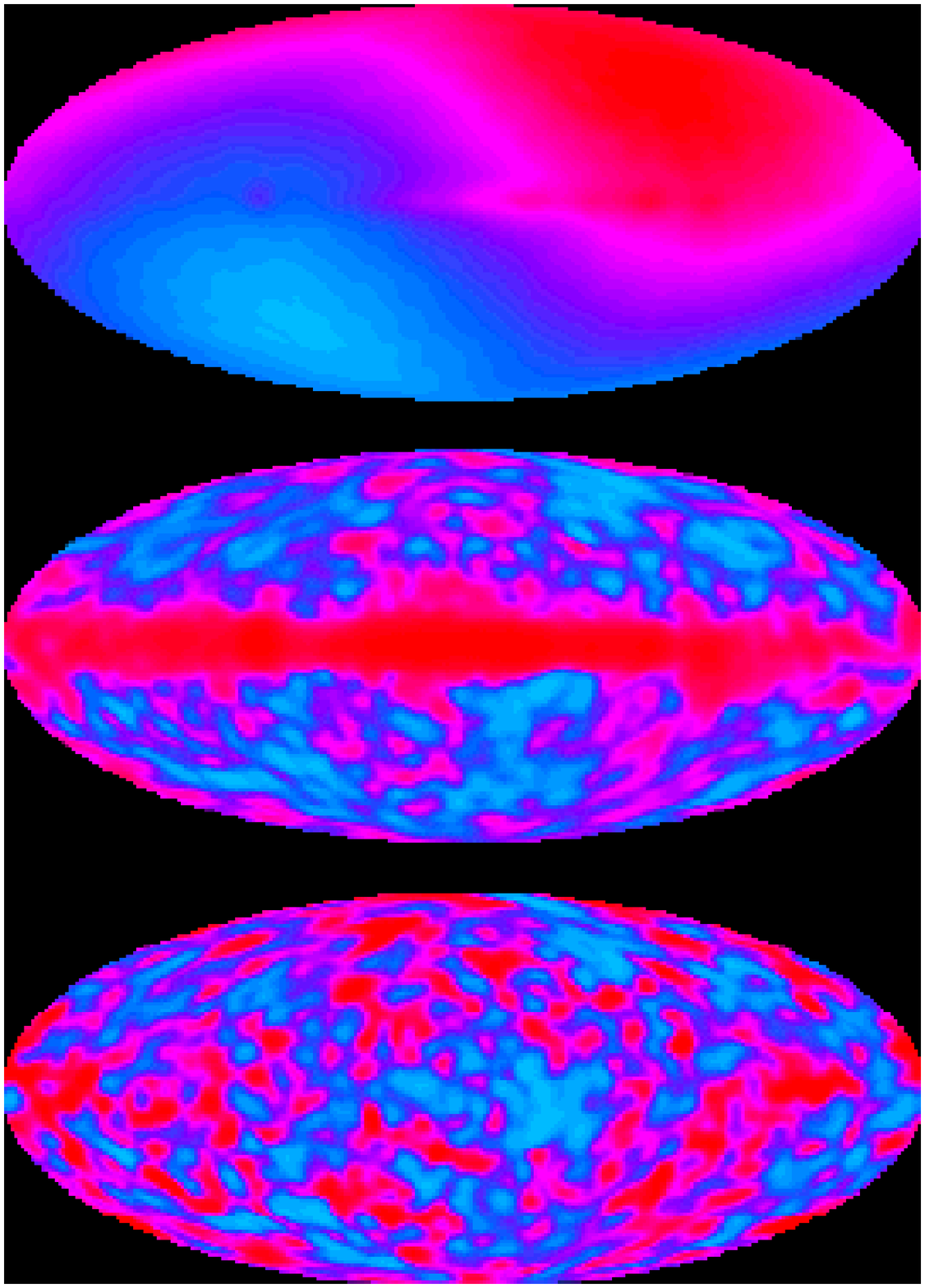} \caption{Sky maps of the temperature fluctuations in the CMB as observed by the NASA satellite mission COBE. The sky map on the top shows the dipole anisotropy after the mean background temperature of $T_0=2.725\,\text{K}$ has been subtracted. The amplitude of the dipole anisotropy is about $3\,\text{mK}$. Also subtracting the dipole yields the sky map in the middle. One sees the small temperature fluctuations whose amplitude is roughly $30\,\mu\text{K}$. But one also sees a lot of foreground contamination along the equator that comes from nearby stars in our galaxy. After removing the foregrounds one finally gets the sky map on the bottom showing the temperature fluctuations in the CMB. Downloaded from \cite{WMAP2003}.} \label{fig:9.1} \end{figure}
\begin{figure} \centering \includegraphics{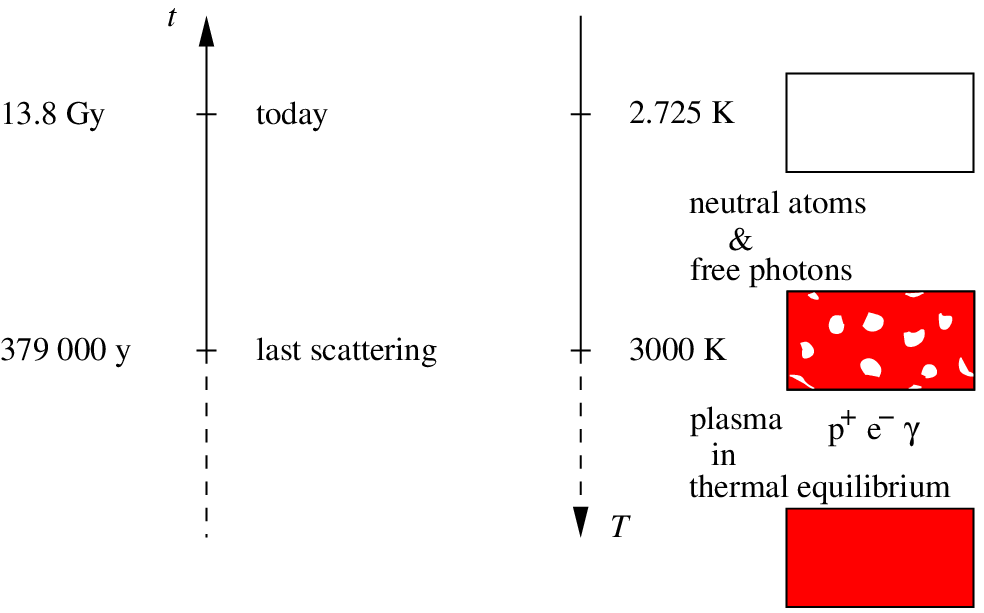} \caption{The expanding universe. At the time of last scattering occured a phase transition from an opaque to a transparent universe.} \label{fig:9.2} \end{figure} \par
The CMB is a relic from the primeval fireball of the early universe. It is the light that comes from the time when the universe was $379\,000$ years old \cite{BennettETal2003}. It was predicted by Gamow in 1948 and explained in detail by Peebles \cite{Peebles1965}. In 1978, Penzias and Wilson \cite{PenziasWilson1965} won the Nobel Prize for Physics for first measuring the CMB at a wavelength of $7.35\,\text{cm}$. Within the resolution of their experiment they found the CMB to be completely isotropic over the whole sky. Later with the much better resolution of the NASA satellite mission Cosmic Background Explorer (COBE), Smoot et al. \cite{SmootETal1992} found fluctuations in the CMB which are of amplitude $10^{-5}$ relative to the mean background temperature of $T_0=2.725\,\text{K}$, except for the large dipole moment, see figure \ref{fig:9.1}. These small fluctuations serve as a fingerprint of the early universe, since the temperature fluctuations are related to the density fluctuations at the time of last scattering. They show how isotropic the universe was at early times. In the inflationary scenario the fluctuations originate from quantum fluctuations which are inflated to macroscopic scales. Due to gravitational instabilities the fluctuations grow steadily and give rise to the formation of stars and galaxies. \par
The theoretical framework in which the CMB and its fluctuations are explained is Einstein's general theory of relativity \cite{Einstein1915a,Einstein1915b,Einstein1915c,Einstein1916,Einstein1917}. Thereby a homogeneous and isotropic background given by a Ro\-bert\-son-Wal\-ker universe \cite{Friedmann1922,Friedmann1924,Lemaitre1927} is perturbed. The time-evolution of the perturbations can be computed in the framework of linear perturbation theory \cite{Lifshitz1946,Bardeen1980}. \par
An explanation for the presence of the CMB is the following, see also figure \ref{fig:9.2}: We live in an expanding universe \cite{Hubble1929}. At early enough times the universe was so hot and dense that it was filled with a hot plasma consisting of ionised atoms, unbounded electrons, and photons. Due to Thomson scattering of photons with electrons, the hot plasma was in thermal equilibrium and the mean free path of the photons was small, hence the universe was opaque. Due to its expansion the universe cooled down and became less dense. When the universe was around $379\,000$ years old, its temperature $T$ has dropped down to approximately $3000\,\text{K}$. At this time, called the time of last scattering, the electrons got bound to the nuclei forming a gas of neutral atoms, mainly hydrogen and helium, and the universe became transparent. Since this time the photons travel freely on their geodesics through the universe. At the time of last scattering the photons had an energy distribution according to a Planck spectrum with temperature of nearly $3000\,\text{K}$. The further expansion of the universe redshifted the photons such that they nowadays have an energy distribution according to a Planck spectrum with temperature of $T_0=2.725\,\text{K}$. This is what we observe as the CMB. \par
Due to the thermal equilibrium before the time of last scattering the CMB is nearly perfectly isotropic, but small density fluctuations lead to small temperature fluctuations. The reason for the small temperature fluctuations comes from a variety of effects. The most dominant effects are the gravitational redshift that is larger in the directions of overdense regions, the intrinsic temperature fluctuations, and the Doppler effect due to the velocity of the plasma.

\section{Ro\-bert\-son-Wal\-ker universes} \label{sec:10} Assuming a universe whose spatial part is locally homogeneous and isotropic, its metric is given by the Ro\-bert\-son-Wal\-ker metric,
\begin{align} ds^2=dt^2-\tilde{a}^2(t)\gamma_{ij}dx^{i}dx^{j}, \label{eq:10.1} \end{align}
where we use the Einstein summation convention. Notice that we have changed the notation slightly. Instead of the quaternion $z$ for the spatial variables, we now write $x=x_0+\ii x_1+\ii x_2$. $\gamma_{ij}$ is the metric of a homogeneous and isotropic three-dimensional space, and the units are rescaled such that the speed of light is $c=1$. Introducing the conformal time $d\eta=\frac{dt}{\tilde{a}(t)}$ we have
\begin{align} ds^2=a^2(\eta)\big[d\eta^2-\gamma_{ij}dx^{i}dx^{j}\big], \label{eq:10.2} \end{align}
where $a(\eta)=\tilde{a}(t(\eta))$ is the cosmic scale factor. \par
With the Ro\-bert\-son-Wal\-ker metric the Einstein equations simplify to the Friedmann equations \cite{Friedmann1922,Friedmann1924,Lemaitre1927}. One of the two Friedmann equations reads
\begin{align} a'^2+\kappa a^2=\frac{8\pi G}{3}T_{0}^{0}a^4+\frac{1}{3}\Lambda a^4 \label{eq:10.3}, \end{align}
and the other Friedmann equation is equivalent to local energy conservation. $a'$ is the derivative of the cosmic scale factor with respect to the conformal time $\eta$. $\kappa$ is the curvature parameter which we choose to be negative, $\kappa=-1$. $G$ is Newton's gravitational constant, $T^{\mu}_{\nu}$ is the energy-momentum tensor, and $\Lambda$ is the cosmological constant. \par
Assuming the energy and matter in the universe to be a perfect fluid consisting of radiation, non-relativistic matter, and a cosmological constant, the time-time component of the energy-momentum tensor reads
\begin{align} T^{0}_{0}=\varepsilon_{\text{r}}(\eta)+\varepsilon_{\text{m}}(\eta), \label{eq:10.4} \end{align}
where the energy densities of radiation and matter scale like
\begin{align} \varepsilon_{\text{r}}(\eta)=\varepsilon_{\text{r}}(\eta_0)\big(\frac{a(\eta_0)}{a(\eta)}\big)^4 \quad \text{and} \quad \varepsilon_{\text{m}}(\eta)=\varepsilon_{\text{m}}(\eta_0)\big(\frac{a(\eta_0)}{a(\eta)}\big)^3. \label{eq:10.5} \end{align}
Here $\eta_0$ denotes the conformal time at the present epoch. \par
Specifying the initial conditions (Big Bang!) $a(0)=0,\ a'(0)>0$, the Friedmann equation (\ref{eq:10.3}) can be solved analytically \cite{AurichSteiner2001},
\begin{align} a(\eta)=\frac{-\big(\frac{\Omega_{\text{r}}}{\Omega_{\text{c}}}\big)^{\frac{1}{2}}{\cal P}'(\eta)+\frac{1}{2}\big(\frac{\Omega_{\text{m}}}{\Omega_{\text{c}}}\big)\big({\cal P}(\eta)-\frac{1}{12}\big)}{2\big({\cal P}(\eta)-\frac{1}{12}\big)^2-\frac{1}{2}\frac{\Omega_{\Lambda}\Omega_{\text{r}}}{\Omega_{\text{c}}^2}}a(\eta_0), \label{eq:10.6} \end{align}
where ${\cal P}(\eta)$ denotes the Weierstrass ${\cal P}$-function which can numerically be evaluated very efficiently, see \cite{AbramowitzStegun1964}, by
\begin{align} {\cal P}(\eta)={\cal P}(\eta;g_2,g_3)=\frac{1}{\eta^2}+\sum_{n=2}^{\infty}c_n\eta^{2n-2} \label{eq:10.7} \end{align}
with
\begin{align} c_2=\frac{g_2}{20}, \quad c_3=\frac{g_3}{28}, \quad \text{and} \quad c_n=\frac{3}{(2n+1)(n-3)}\sum_{m=2}^{n-2}c_mc_{n-m} \quad \text{for $n\ge4$}
. \label{eq:10.8} \end{align}
The so-called invariants $g_2$ and $g_3$ are determined by the cosmological parameters,
\begin{align} g_2=\frac{\Omega_{\Lambda}\Omega_{\text{r}}}{\Omega_{\text{c}}^2}+\frac{1}{12}, \quad g_3=\frac{1}{6}\frac{\Omega_{\Lambda}\Omega_{\text{r}}}{\Omega_{\text{c}}^2}-\frac{1}{16}\frac{\Omega_{\Lambda}\Omega_{\text{m}}^2}{\Omega_{\text{c}}^3}-\frac{1}{216}, \label{eq:10.9} \end{align}
with
\begin{align}
&\Omega_{\text{r}}=\frac{8\pi G\varepsilon_{\text{r}}(\eta_0)}{3H^2(\eta_0)}, \label{eq:10.10} \\
&\Omega_{\text{m}}=\frac{8\pi G\varepsilon_{\text{m}}(\eta_0)}{3H^2(\eta_0)}, \label{eq:10.11} \\
&\Omega_{\text{c}}=\frac{1}{H^2(\eta_0)a^2(\eta_0)} \label{eq:10.12}, \\
&\Omega_{\Lambda}=\frac{\Lambda}{3H^2(\eta_0)}, \label{eq:10.13}
\end{align}
where
\begin{align} H(\eta)=\frac{a'(\eta)}{a^2(\eta)} \label{eq:10.14} \end{align}
is the Hubble parameter. \par
Because of the homogeneity and isotropy, nowhere in the equations of a Ro\-bert\-son-Wal\-ker universe appears the Laplacian.

\section{Perturbed Ro\-bert\-son-Wal\-ker universes} \label{sec:11} The idealisation to an exact homogeneous and isotropic universe was essential to derive the spacetime of the Ro\-bert\-son-Wal\-ker universe. But obviously, we do not live in a universe which is perfectly homogeneous and isotropic. We see individual stars, galaxies, and in between large empty space. Knowing the spacetime of the Ro\-bert\-son-Wal\-ker universe, we can study small perturbations around the homogeneous and isotropic background. Since the amplitude of the large scale fluctuations in the universe is of relative size $10^{-5}$ \cite{SmootETal1992}, we can use linear perturbation theory. In longitudinal gauge the most general scalar perturbation of the Ro\-bert\-son-Wal\-ker metric reads
\begin{align} ds^2=a^2(\eta)\big[(1+2\Phi)d\eta^2-(1-2\Psi)\gamma_{ij}dx^{i}dx^{j}\big], \label{eq:11.1} \end{align}
where $\Phi=\Phi(\eta,x)$ and $\Psi=\Psi(\eta,x)$ are functions of spacetime. \par
Assuming that the energy and matter density in the universe can be described by a perfect fluid, consisting of radiation, non-relativistic matter, and a cosmological constant, and neglecting possible entropy perturbations, the Einstein equations reduce in first order perturbation theory \cite{MukhanovFeldmanBrandenberger1992} to
\begin{align} &\Phi=\Psi, \label{eq:11.2} \\
&\Phi''+3\hat{H}(1+c_{\text{s}}^2)\Phi'-c_{\text{s}}^2\Delta\Phi+\big(2\hat{H}'+(1+3c_{\text{s}}^2)(\hat{H}^2+1)\big)\Phi=0, \label{eq:11.3} \end{align}
where $\hat{H}=\frac{a'}{a}$ and $c_{\text{s}}^2=(3+\frac{9}{4}\frac{\varepsilon_{\text{m}}}{\varepsilon_{\text{r}}})^{-1}$ are given by the solution of the non-perturbed Ro\-bert\-son-Wal\-ker universe. In the partial differential equation (\ref{eq:11.3}) the Laplacian occurs. If the initial and the boundary conditions of $\Phi$ are specified, the time-evolution of the metric perturbations can be computed. \par
With the separation ansatz
\begin{align} \Phi(\eta,x)=\sum_{k}f_k(\eta)\psi_k(x)+\int dk\,f_k(\eta)
\psi_k(x), \label{eq:11.4} \end{align}
where the $\psi_k$ are the eigenfunctions of the negative Laplacian, and the $E_k$ are the corresponding eigenvalues,
\begin{align} -\Delta\psi_k(x)=E_k\psi_k(x), \label{eq:11.5} \end{align}
(\ref{eq:11.3}) simplifies to
\begin{align} f_k''(\eta)+3\hat{H}(1+c_{\text{s}}^2)f_k'(\eta)+\big(c_{\text{s}}^2E_k+2\hat{H}'+(1+3c_{\text{s}}^2)(\hat{H}^2+1)\big)f_k(\eta)=0. \label{eq:11.6} \end{align}
The ODE (\ref{eq:11.6}) can be computed numerically in a straightforward way, and we finally obtain the metric of the whole universe. This gives the input to the Sachs-Wolfe formula \cite{SachsWolfe1967} which connects the metric perturbations with the temperature fluctuations,
\begin{align} \frac{\delta T}{T_0}(\hat{n})=2\Phi(\eta_{\text{SLS}},x(\eta_{\text{SLS}}))-\frac{3}{2}\Phi(0,x(0))+2\int_{\eta_{\text{SLS}}}^{\eta_0}d\eta\,\frac{\partial}{\partial\eta}\Phi(\eta,x(\eta)), \label{eq:11.7} \end{align}
where $\hat{n}$ is a unit vector in the direction of the observed photons. $x(\eta)$ is the geodesic along which the light travels from the surface of last scattering (SLS) towards us, and $\eta_{\text{SLS}}$ is the time of last scattering. \par
If we choose the topology of the universe to be the orbifold of the Picard group, we can use in (\ref{eq:11.4}) the Maass cusp forms and the Eisenstein series computed in sections \ref{sec:6}--\ref{sec:8a}. Let us further choose the initial conditions to be
\begin{align} f_k(0)=\frac{\sigma_k\alpha}{\sqrt{E_k\sqrt{E_k-1}}} \quad \text{and} \quad f_k'(0)=\frac{-\Omega_{\text{m}}f_k(0)}{16(\Omega_{\text{c}}\Omega_{\text{r}})^{\frac{1}{2}}}, \label{eq:11.8} \end{align}
\cite{LevinBarrowBunnSilk1997,InoueTomitaSugiyama2000,AurichSteiner2001,Levin2002}, which carry over to a Har\-ri\-son-Zel'\-do\-vich spectrum having a spectral index $n=1$ and selecting only the non-decaying modes. $\alpha$ is a constant independent of $k$ which is fitted to the amplitude of the observed temperature fluctuations. The quantities $\sigma_k$ are random signs, $\sigma_k\in\{-1,1\}$. \par
\begin{figure} \centering \includegraphics[height=11cm,angle=-90]{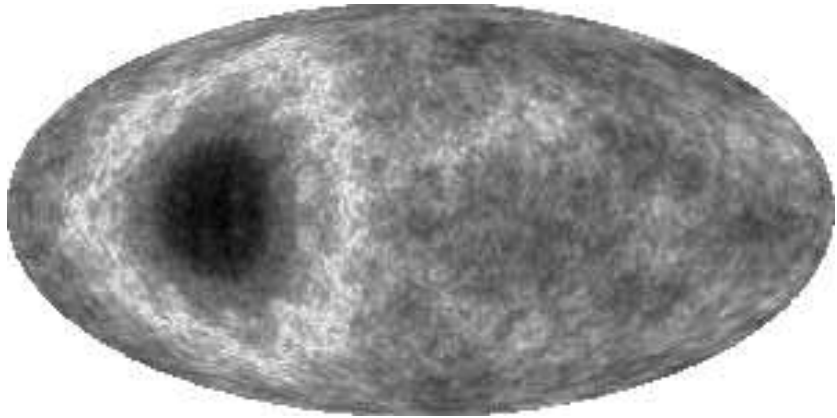} \caption{The sky map of the calculated temperature fluctuations of the CMB for $\Omega_{\text{tot}}=0.9,\Omega_{\text{m}}=0.3,\Omega_{\Lambda}=0.6,h_0=0.65$, and $x_{\text{obs}}=0.2+0.1\ii+1.6\ij$, where the Eisenstein series and the cusp forms are taken into account. The coordinate system is oriented such that the cusp is located in the left part of the figure where the hot spot can be seen.} \label{fig:11.1} \end{figure}
\begin{figure} \centering \includegraphics[height=11cm,angle=-90]{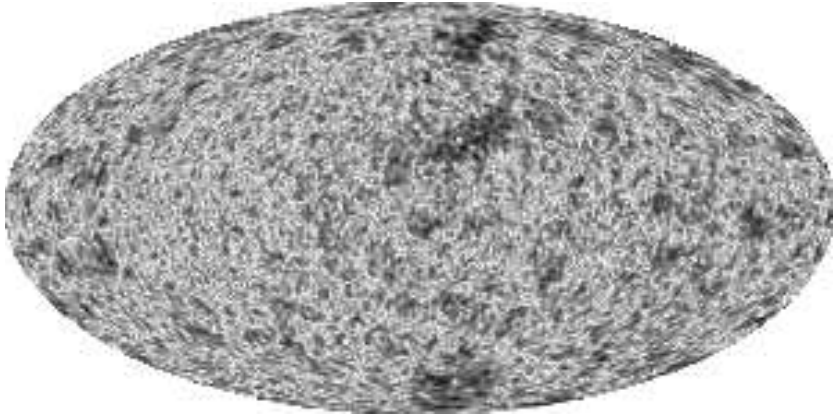} \caption{The sky map of the calculated temperature fluctuations of the CMB for $\Omega_{\text{tot}}=0.9,\Omega_{\text{m}}=0.3,\Omega_{\Lambda}=0.6,h_0=0.65$, and $x_{\text{obs}}=0.2+0.1\ii+1.6\ij$, if only the cusp forms are taken into account.} \label{fig:11.2} \end{figure}
\begin{figure} \centering \includegraphics[width=11cm]{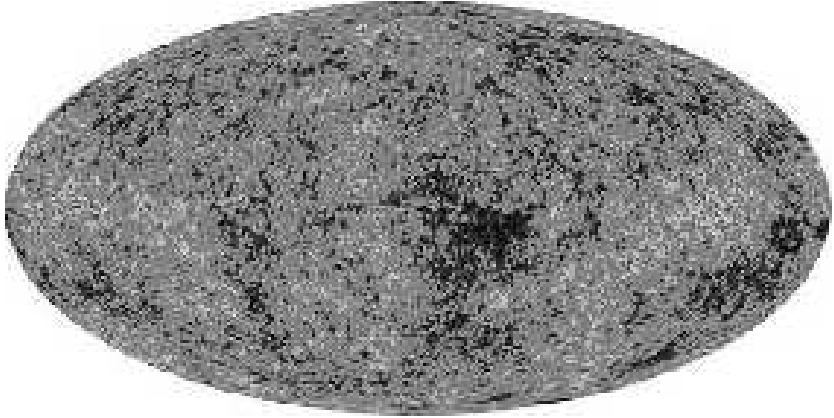} \caption{The temperature fluctuations of the CMB observed by WMAP. Downloaded from \cite{WMAP2003}.} \label{fig:11.3} \end{figure}

The following cosmological parameters are used $\Omega_{\text{m}}=0.3,\Omega_{\Lambda}=0.6,\Omega_{\text{c}}=1-\Omega_{\text{tot}}=1-\Omega_{\text{r}}-\Omega_{\text{m}}-\Omega_{\Lambda},H(\eta_0)=100\,h_0\,\text{km}\,\text{s}^{-1}\,\text{Mpc}^{-1}$ with $h_0=0.65$. The density $\Omega_{\text{r}}\approx10^{-4}$ is determined by the current temperature $T_0=2.725\,\text{K}$. The point of the observer is chosen to be at $x_{\text{obs}}=0.2+0.1\ii+1.6\ij$. The sky map of figure \ref{fig:11.1} is computed with the expansion (\ref{eq:11.4}) using cusp forms and Eisenstein series. For numerical reasons the infinite spectrum is cut such that only the eigenvalues with $E=k^2+1\le19601$ and their corresponding eigenfunctions are taken into account. In addition the integration of the continuous spectrum is approximated numerically using a Gauss quadrature with $16$ integration points per unit interval. The resulting map in figure \ref{fig:11.1} is not in agreement with the cosmological observations \cite{BennettETal2003} which are shown in figure \ref{fig:11.3}. Especially in the direction of the cusp, the temperature fluctuations of our calculated model show a strong peak. Clearly, such a pronounced hot spot is unphysical, since it is not observed. This peak comes from the contribution of the Eisenstein series. The sky map in figure \ref{fig:11.2}, which results from taking only the cusp forms into account, yields a much better agreement with the cosmological observations.
 \par
The hot spot in the direction of the cusp, which results from the contribution of the Eisenstein series, is sometimes used, see e.g.\ \cite{SokolovStarobinskii1976}, as an argument that the Eisenstein series should not be taken into account. Some papers even ignore the existence of the Eisenstein series completely, e.g.\ \cite{LevinBarrowBunnSilk1997}. \par
Another possibility is to take the Eisenstein series into account, but to choose for them initial conditions which differ from (\ref{eq:11.8}) such that the intensity towards the cusp is not increasing. To see whether this is possible, let us consider the completeness relation of Maass waveforms,
\begin{align} \Phi(\eta,x)=\sum_{n\ge0}\langle\psi_{k_n},\Phi(\eta,\cdot)\rangle\psi_{k_n}(x)+\int_{0}^{\infty}\langle\psi^{\text{Eisen}}_k,\Phi(\eta,\cdot)\rangle\psi^{\text{Eisen}}_k(x)\,dk, \label{eq:11.9} \end{align}
cf. (\ref{eq:6.12}) and (\ref{eq:6.14}), saying that within the fundamental cell any desired metric perturbation $\Phi(\eta_{\text{SLS}},x)$ at the time of last scattering can be expressed via the conditions
\begin{align*} &f_{k_n}(\eta_{\text{SLS}})=\langle\psi_{k_n},\Phi(\eta_{\text{SLS}},\cdot)\rangle \quad &&\text{for the discrete spectrum}, \\ &f^{\text{Eisen}}_k(\eta_{\text{SLS}})=\langle\psi^{\text{Eisen}}_k,\Phi(\eta_{\text{SLS}},\cdot)\rangle \quad &&\text{for the continuous spectrum}. \end{align*}
E.g., choosing the metric perturbation to be
\begin{align} \Phi(\eta_{\text{SLS}},x)=\cos(k'\ln x_2) \label{eq:11.10} \end{align}
would neither result in a hot spot in the direction of the cusp nor would $f^{\text{Eisen}}_k(\eta)$ vanish identically. A more general ansatz would be
\begin{align} \Phi(\eta_{\text{SLS}},x)=\int_{0}^{\infty}dk'\,\big(A_{k'}(\eta_{\text{SLS}},x)\cos(k'\ln x_2)+B_{k'}(\eta_{\text{SLS}},x)\sin(k'\ln x_2)\big), \label{eq:11.11} \end{align}
where $A_{k'}(\eta_{\text{SLS}},x)$ and $B_{k'}(\eta_{\text{SLS}},x)$ are some amplitudes. \par
Concerning the initial conditions we found that ({\ref{eq:11.8}) has to be modified for the continuous part of the spectrum. Hence, the important question occurs: How do the correct initial conditions look like? Unfortunately, this question has not been answered yet. Work in progress is concerned with finding the correct initial conditions and their physical motivation \cite{AurichSteinerThen2004}. \par
Not knowing the correct initial conditions, we use ({\ref{eq:11.8}) for the discrete part and neglect the continuous part of the spectrum. We proceed with the discussion of the properties that come from the discrete part of the spectrum, only.

\section{Comparison with the cosmological observations} \label{sec:12}
\begin{figure} \centering \psfrag{l}{\small $l$} \psfrag{l(l+1)Cl/2pi}{$\frac{l(l+1)C_l}{2\pi}$} \psfrag{muK2}{\scriptsize in $\mu K^2$} \psfrag{ 0.30103}{\scriptsize $\ \ \ \,2$} \psfrag{ 1.30103}{\scriptsize $\ \ \,20$} \psfrag{ 2.30103}{\scriptsize $\ \ 200$} \includegraphics{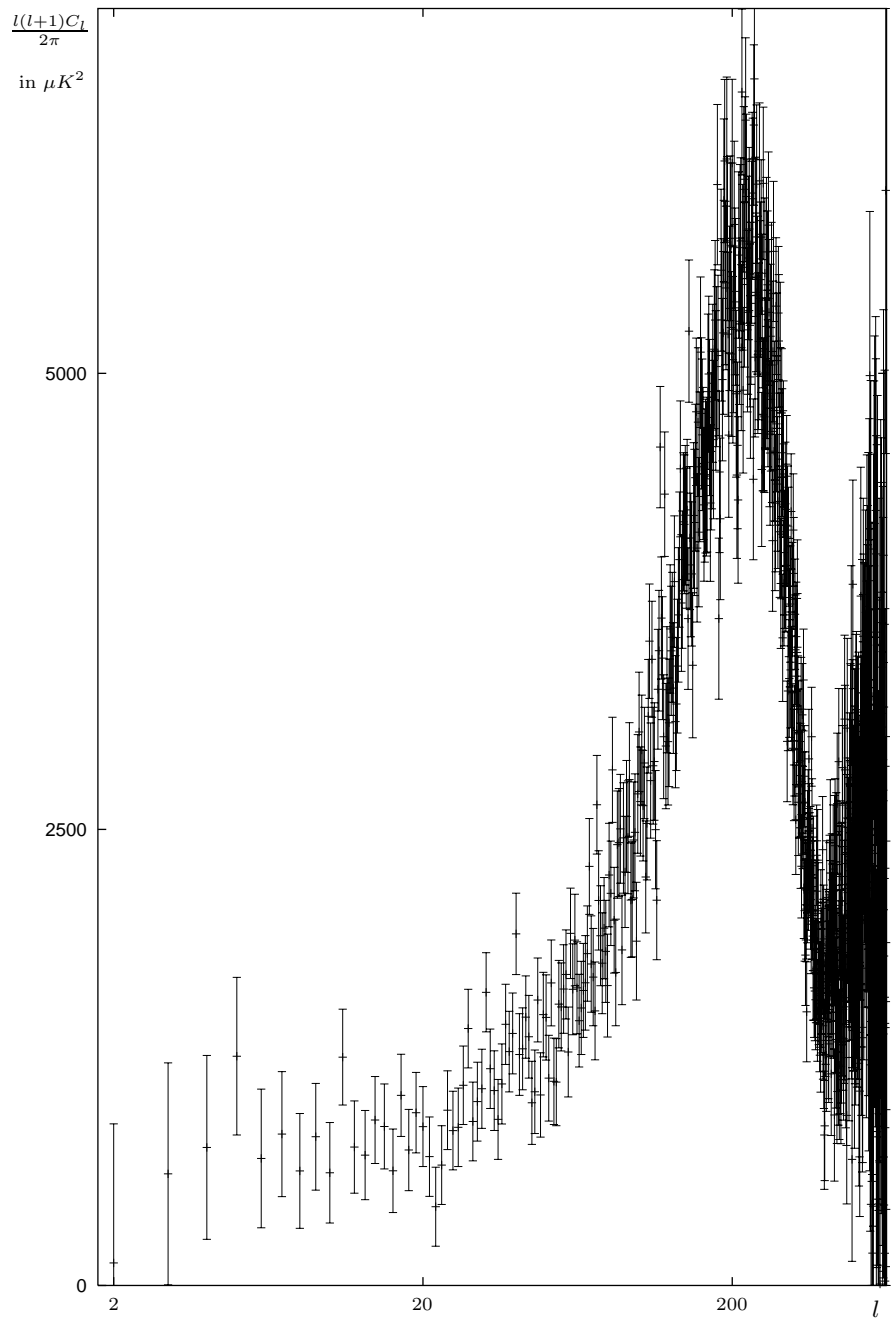} \caption{The angular power spectrum of the temperature fluctuations observed by WMAP. Downloaded from \cite{WMAP2003}.} \label{fig:12.1} \end{figure}
The key experiments that observed the fluctuations of the CMB are: COBE, Smoot et al. \cite{SmootETal1992}; Boomerang, de Bernardis et al. \cite{BernardisETal2000}; MA\-XI\-MA-1, Hanany et al. \cite{HananyETal2000}; ACBAR, Kuo et al. \cite{KuoETal2002}; CBI, Pearson et al. \cite{PearsonETal2003}; WMAP, Bennett et al. \cite{BennettETal2003}. \par
Concerning the topology of the universe which manifests itself in the low multipole moments, there exist only the cosmological observations from COBE and WMAP. These two experiments have scanned a full sky map of the temperature fluctuations. The NASA satellite mission of the Cosmic Background Explorer (COBE) was the first experiment that detected the temperature fluctuations of the CMB in 1992. In addition, it found that the quadrupole moment of the temperature fluctuations is suppressed. In 2003 the NASA satellite mission of the Wilkinson Microwave Anisotropy Probe (WMAP) scanned a full sky map of the CMB with a much higher resolution confirming and improving the results of COBE. All the other experiments were either ground based or balloon flights that could only scan parts of the sky and thus did not measure the first multipole moments. Their strength was to measure the finer-scale anisotropy manifested in the higher multipole moments. \par
In order to quantitatively compare our results with the observations, we introduce the angular power spectrum and the correlation function. Expanding the temperature fluctuations in the CMB into spherical harmonics,
\begin{align} \delta T\big(\hat{n}(\theta,\phi)\big)=\sum_{l=2}^{\infty}\sum_{m=-l}^{l}a_{lm}Y_{l}^{m}(\theta,\phi), \label{eq:12.1a} \end{align}
yields the expansion coefficients $a_{lm}$. We remark that in (\ref{eq:12.1a}) the sum over $l$ starts at $l=2$, i.e.\ the monopole and dipole have been subtracted, because the monopole term does not give rise to an anisotropy and the dipole term is unobservable due to the peculiar velocity of our earth relative to the background. From the expansion coefficients we obtain the multipole moments of the CMB anisotropies,
\begin{align} C_l=\frac{1}{2l+1}\sum_{m=-l}^{l}|a_{lm}|^2, \label{eq:12.2a} \end{align}
the angular power spectrum,
\begin{align} \frac{l(l+1)}{2\pi}C_l, \label{eq:12.3a} \end{align}
and the two-point correlation function,
\begin{align} C(\vartheta)=\big\langle\delta T(\hat{n})\delta T(\hat{n}')\big\rangle_{\cos\vartheta=\hat{n}\cdot\hat{n}'}=\frac{1}{8\pi^2\sin\vartheta}\int\int_{\substack{S^2\times S^2 \\ \hat{n}\cdot\hat{n}'=\cos\vartheta}}d\hat{n}\,d\hat{n}'\,\delta T(\hat{n})\delta T(\hat{n}'). \label{eq:12.4a} \end{align}
Assuming a Gauss\-ian value distribution for the expansion coefficients $a_{lm}$, the two-point correlation function is related to the multipole moments via
\begin{align} C(\vartheta)=\frac{1}{4\pi}\sum_{l=2}^{\infty}(2l+1)C_lP_l(\cos\vartheta). \label{eq:12.5a} \end{align}
When determining the correlation function numerically, we consider (\ref{eq:12.5a}) as its definition irrespectively whether the expansion coefficients $a_{lm}$ are Gauss\-ian distributed or not. \par
Figures \ref{fig:12.1}--\ref{fig:12.4} show the angular power spectrum and the correlation function, respectively, measured by WMAP \cite{SpergelETal2003}. In figure \ref{fig:12.1} we see that the first multipole moments resulting from the cosmological observations are suppressed, especially the quadrupole, $C_2$. The next few multipole moments up to $l\approx25$ give rise to a plateau in the angular power spectrum, before the larger multipole moments with $l>25$ start to increase until they reach the first acoustic peak. Page et al. \cite{PageETal2003} find that the first acoustic peak is at $l=220.1\pm0.8$. The trough following this peak is at $l=411.7\pm3.5$, and the second peak is at $l=546\pm10$. \par
\begin{figure} \centering \psfrag{l}{\small $l$} \psfrag{l(l+1)Cl/2pi}{$\frac{l(l+1)C_l}{2\pi}$} \psfrag{muK2}{\scriptsize in $\mu K^2$} \psfrag{ 0.30103}{\scriptsize $\ \ \ \,2$} \psfrag{ 1.30103}{\scriptsize $\ \ \,20$} \psfrag{ 2.30103}{\scriptsize $\ \ 200$} \includegraphics{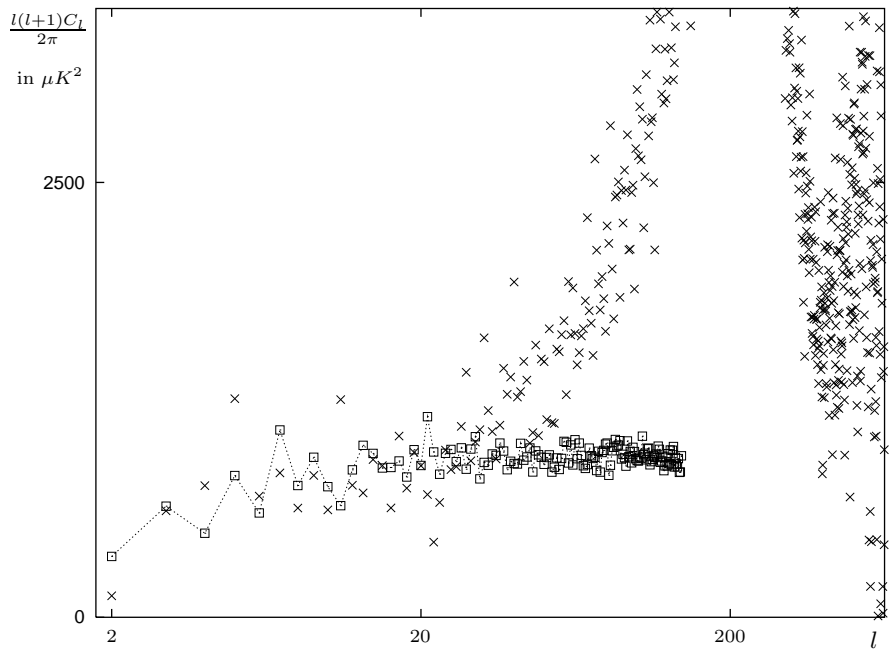} \caption{The angular power spectrum of the calculated temperature fluctuations for $\Omega_{\text{tot}}=0.9,\Omega_{\text{m}}=0.3,\Omega_{\Lambda}=0.6,h_0=0.65$, and $x_{\text{obs}}=0.2+0.1\ii+1.6\ij$ $(\Box)$ in comparison with the angular power spectrum of the WMAP observations $(\times)$.} \label{fig:12.3} \end{figure}
Figures \ref{fig:12.3} and \ref{fig:12.4} show the angular power spectrum and the correlation function, respectively, corresponding to the calculated sky map for the Picard model of figure \ref{fig:11.2} in comparison with the results of the cosmological observations and with the concordance model \cite{SpergelETal2003}. Regarding our results of the calculated temperature fluctuations, see figure \ref{fig:12.3}, we see that the first multipole moments are suppressed in accordance with the cosmological observations. Especially the quadrupole term, $C_2$, computed in our model, is suppressed. However the suppression of the observed one is still stronger. Note, that in the case of the concordance model, the first multipoles are increased in contrast to our results. This points to a non-trivial topology of our universe. \par
\begin{figure} \centering \psfrag{theta}{\small $\vartheta$} \psfrag{C(theta)}{\small $C(\vartheta)$} \psfrag{muK2}{\scriptsize in $\mu K^2$} \includegraphics{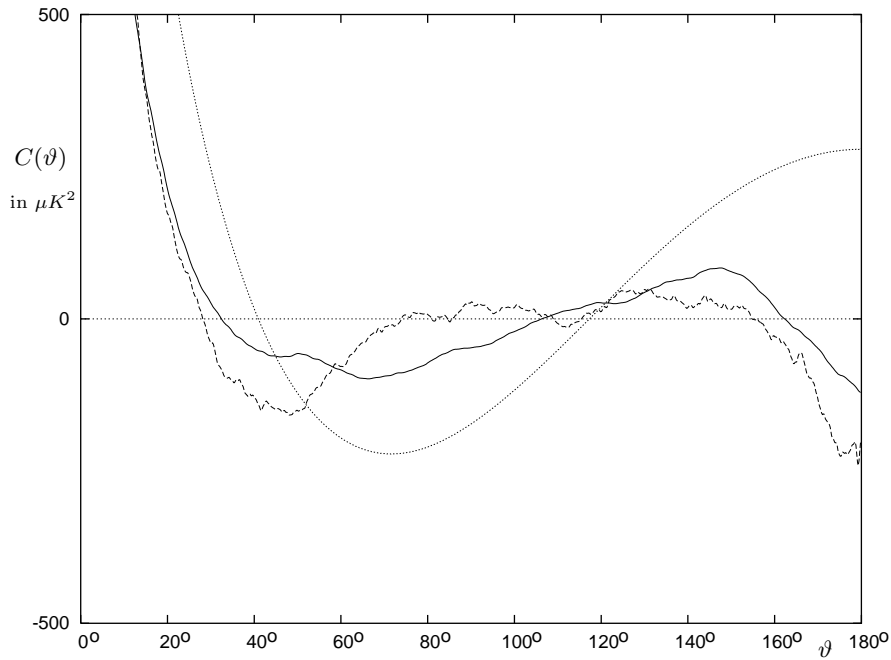} \caption{The solid line is the correlation function of the calculated temperature fluctuations for $\Omega_{\text{tot}}=0.9,\Omega_{\text{m}}=0.3,\Omega_{\Lambda}=0.6,h_0=0.65$, and $x_{\text{obs}}=0.2+0.1\ii+1.6\ij$. The dashed line represents the correlation function of the WMAP observations, and the dotted line shows the concordance model \cite{SpergelETal2003}.} \label{fig:12.4} \end{figure}
The next multipole moments give rise to a plateau in the angular power spectrum. We use this plateau to fit the constant $\alpha$ of the initial conditions (\ref{eq:11.8}) such that the sum of the first $19$ computed multipole moments matches with the cosmological observations,
\begin{align} \sum_{l=2}^{20}C_{l,\text{computed}}\overset{!}{=}\sum_{l=2}^{20}C_{l,\text{observed}}. \label{eq:12.1} \end{align}
But concerning the fluctuations on finer scales (smaller angular resolution), i.e.\ higher multipoles, we do not obtain the acoustic peaks in our numerical calculations. This is due to the fact that we neglect several physical effects which perturb the temperature on finer scales. We ignore all the physical effects that dominate on scales smaller than those corresponding to $l\gg25$, since our main interest is in the influence of the non-trivial topology of the universe due to the Picard group. \par
Another comparison of our calculated temperature fluctuations with the cosmological observations can be done via the correlation function (\ref{eq:12.5a}) which emphasises large scales, i.e.\ the behaviour of low multipoles. Concerning the correlation function $C(\vartheta)$, we find quite good agreement of our calculated temperature fluctuations with the cosmological observations, see figure \ref{fig:12.4}. The reason for this good agreement of our model with the observations is due to the suppressed quadrupole moment. We also show in figure \ref{fig:12.4} (dotted line) the concordance model \cite{SpergelETal2003} which is not in good agreement with the data for $\vartheta\gtrsim7^{\circ}$ in contrast to our model (full curve). Especially for large angular separations, $\vartheta\gtrsim160^{\circ}$, the concordance model is not able to describe the observed anticorrelation in $C(\vartheta)$. This anticorrelation constitutes a fingerprint in the CMB that favours a non-trivial topology for the universe.

\section{Conclusion} \label{sec:13} We pointed out the importance of the eigenvalue equation of the Laplacian and gave some physical applications. In the framework of quantum chaos, we focused on the eigenvalue equation of the Picard orbifold and computed the solutions numerically. Our main goal was to determine the statistical properties of the solutions, but also to apply the solutions to a completely different problem, namely the temperature fluctuations in the CMB. Here, we demonstrate that a model of the universe with the topology of the Picard group matches the large-scale anisotropy in the CMB better than the current concordance model. Thereby, we show that the methods developed in quantum chaos can be successfully applied to other fields of research although the physical interpretation can differ completely.

\section{Acknowledgments} \label{sec:14} H.~T. greatfully acknowledges the encouraging advice of Prof. Dennis A. Hejhal. This work has been supported by the European Commission under the Research Training Network (Mathematical Aspects of Quantum Chaos) no HPRN-CT-2000-00103 of the IHP Programme and by the Deutsche For\-schungs\-ge\-mein\-schaft under the contract no. DFG Ste 241/16-1. The computations were run on the computers of the Universit\"{a}ts-Rechenzentrum Ulm.

\newcommand{\etalchar}[1]{$^{#1}$}

\printindex

\begin{thebibliography}{dBAB{\etalchar{+}}00}

\bibitem[ABS91]{AurichBogomolnySteiner1991}
R.~Aurich, E.~B. Bogomolny, and F.~Steiner.
\newblock Periodic orbits on the regular hyperbolic octagon.
\newblock {\em Physica D}, 48:91--101, 1991.

\bibitem[AS64]{AbramowitzStegun1964}
M.~Abramowitz and I.~A. Stegun.
\newblock {\em Handbook of mathematical functions with formulas, graphs, and
  mathematical tables}.
\newblock National Bureau of Standards Applied Mathematics Series, 55. U.S.
  Government Printing Office, Washington, D.C., 1964.

\bibitem[AS88]{AurichSteiner1988}
R.~Aurich and F.~Steiner.
\newblock On the periodic orbits of a strongly chaotic system.
\newblock {\em Physica D}, 32:451--460, 1988.

\bibitem[AS89]{AurichSteiner1989}
R.~Aurich and F.~Steiner.
\newblock Periodic-orbit sum rules for the Hadamard-Gutzwiller model.
\newblock {\em Physica D}, 39:169--193, 1989.

\bibitem[AS90]{AurichSteiner1990}
R.~Aurich and F.~Steiner.
\newblock Energy-level statistics of the Hadamard-Gutzwiller ensemble.
\newblock {\em Physica D}, 43:155--180, 1990.

\bibitem[AS01]{AurichSteiner2001}
R.~Aurich and F.~Steiner.
\newblock The cosmic microwave background for a nearly flat compact hyperbolic
  universe.
\newblock {\em Mon. Not. Roy. Astron. Soc.}, 323:1016--1024, 2001.

\bibitem[ALST04]{AurichSteinerThen2004}
R.~Aurich, S.~Lustig, F.~Steiner, and H.~Then, 2004.
\newblock Hyperbolic universes with a horned topology and the CMB anisotropy.
\newblock {\em astro-ph/0403597}.

\bibitem[Ava56]{Avakumovic1956}
V.~G. Avakumovi\'{c}.
\newblock \"{U}ber die Eigenfunktionen auf geschlossenen Riemannschen
  Mannigfaltigkeiten. (German).
\newblock {\em Math. Z.}, 65:327--344, 1956.

\bibitem[Ave03]{Avelin2002}
H.~Avelin.
\newblock On the deformation of cusp forms (Licentiate Thesis).
\newblock {\em UUDM report 2003:8}, Uppsala 2003.

\bibitem[Bar80]{Bardeen1980}
J.~Bardeen.
\newblock Gauge-invariant cosmological perturbations.
\newblock {\em Phys. Rev. D}, 22:1882--1905, 1980.

\bibitem[Ber77]{Berry1977}
M.~V. Berry.
\newblock Regular and irregular semiclassical wavefunctions.
\newblock {\em J. Phys. A}, 10:2083--2091, 1977.

\bibitem[BGGS92]{BogomolnyGeorgeotGiannoniSchmit1992}
E.~B. Bogomolny, B.~Georgeot, M.-J. Giannoni, and C.~Schmit.
\newblock Chaotic billiards generated by arithmetic groups.
\newblock {\em Phys. Rev. Lett.}, 69:1477--1480, 1992.

\bibitem[BGS84]{BohigasGiannoniSchmit1984}
O.~Bohigas, M.-J. Giannoni, and C.~Schmit.
\newblock Characterization of chaotic quantum spectra and universality of
  level fluctuation laws.
\newblock {\em Phys. Rev. Lett.}, 52:1--4, 1984.

\bibitem[BGS86]{BohigasGiannoniSchmit1986}
O.~Bohigas, M.-J. Giannoni, and C.~Schmit.
\newblock Spectral fluctuations, random matrix theories and chaotic motion.
  Stochastic processes in classical and quantum systems.
\newblock {\em Lecture Notes in Phys.}, 262:118--138, 1986.

\bibitem[BHH{\etalchar{+}}03]{BennettETal2003}
C.~L. Bennett, M.~Halpern, G.~Hinshaw, N.~Jarosik, A.~Kogut, M.~Limon,
  S.~S. Meyer, L.~Page, D.~N. Spergel, G.~S. Tucker, E.~Wollack, E.~L. Wright,
  C.~Barnes, M.~R. Greason, R.~S. Hill, E.~Komatsu, M.~R. Nolta, N.~Odegard,
  H.~V. Peirs, L.~Verde, and J.~L. Weiland.
\newblock First year Wilkinson microwave anisotropy probe (WMAP)
  observations: Preliminary maps and basic results.
\newblock {\em Astroph. J. Suppl.}, 148:1--27, 2003.

\bibitem[Bol93]{Bolte1993}
J.~Bolte.
\newblock Some studies on arithmetical chaos in classical and quantum
  mechanics.
\newblock {\em Int. J. Mod. Phys. B}, 7:4451--4553, 1993.

\bibitem[Bor69]{Borel1969}
A.~Borel.
\newblock {\em Introduction aux groupes arithm\'{e}tiques.} (French).
\newblock Hermann, 1969.

\bibitem[BSS92]{BolteSteilSteiner1992}
J.~Bolte, G.~Steil, and F.~Steiner.
\newblock Arithmetical chaos and violation of universality in energy level
  statistics.
\newblock {\em Phys. Rev. Lett.}, 69:2188--2191, 1992.

\bibitem[BT76]{BerryTabor1976}
M.~V. Berry and M.~Tabor.
\newblock Closed orbits and the regular bound spectrum.
\newblock {\em Proc. Roy. Soc. London Ser. A}, 349:101--123, 1976.

\bibitem[CF86]{ChinburgFriedman1986}
T.~Chinburg and E.~Friedman.
\newblock The smallest arithmetic hyperbolic three-orbifold.
\newblock {\em Invent. Math.}, 86:507--527, 1986.

\bibitem[CFJR01]{ChinburgFriedmanJonesReid2001}
T.~Chinburg, E.~Friedman, K.~N. Jones, and A.~W. Reid.
\newblock The arithmetic hyperbolic 3-manifold of smallest volume.
\newblock {\em Ann. Scuola Norm. Sup. Pisa Cl. Sci.}, 30:1--40, 2001.

\bibitem[dBAB{\etalchar{+}}00]{BernardisETal2000}
P.~de Bernardis, P.~A. R. Ade, J.~J. Bock, J.~R. Bond, J.~Borrill,
  A.~Boscaleri, K.~Coble, B.~P. Crill, G.~De Gasperis, P.~C. Farese,
  P.~G. Ferreira, K.~Ganga, M.~Giacometti, E.~Hivon, V.~V. Hristov,
  A.~Iacoangeli, A.~H. Jaffe, A.~E. Lange, L.~Martinis, S.~Masi, P.~Mason,
  P.~D. Mauskopf, A.~Melchiorri, L.~Miglio, T.~Montroy, C.~B. Netterfield,
  E.~Pascale, F.~Piacentini, D.~Pogosyan, S.~Prunet, S.~Rao, G.~Romeo,
  J.~E. Ruhl, F.~Scaramuzzi, D.~Sforna, and N.~Vittorio.
\newblock A flat universe from high-resolution maps of the cosmic microwave
  background radiation.
\newblock {\em Nature}, 404:955--959, 2000.

\bibitem[Dys70]{Dyson1970}
F.~J. Dyson.
\newblock Correlations between the eigenvalues of a random matrix.
\newblock {\em Commun. Math. Phys.}, 19:235--250, 1970.

\bibitem[EGM85]{ElstrodtGrunewaldMennicke1985}
J.~Elstrodt, F.~Grunewald, and J.~Mennicke.
\newblock Eisenstein series on three-dimensional hyperbolic space and
  imaginary quadratic number fields.
\newblock {\em J. Reine Angew. Math.}, 360:160--213, 1985.

\bibitem[EGM98]{ElstrodtGrunewaldMennicke1998}
J.~Elstrodt, F.~Grunewald, and J.~Mennicke.
\newblock {\em Groups Acting on Hyperbolic Space}.
\newblock Springer, 1998.

\bibitem[Ein15a]{Einstein1915a}
A.~Einstein.
\newblock Zur allgemeinen Relativit\"{a}tstheorie. (German).
\newblock {\em Preu{\ss}. Akad. Wiss., Sitzungsber.}, pp. 778--786, 1915.

\bibitem[Ein15b]{Einstein1915b}
A.~Einstein.
\newblock Zur allgemeinen Relativit\"{a}tstheorie (Nachtrag). (German).
\newblock {\em Preu{\ss}. Akad. Wiss., Sitzungsber.}, pp. 799--801, 1915.

\bibitem[Ein15c]{Einstein1915c}
A.~Einstein.
\newblock Die Feldgleichungen der Gravitation. (German).
\newblock {\em Preu{\ss}. Akad. Wiss., Sitzungsber.}, pp. 844--847, 1915.

\bibitem[Ein16]{Einstein1916}
A.~Einstein.
\newblock Die Grundlage der allgemeinen Relativit\"{a}tstheorie. (German).
\newblock {\em Ann. Phys.}, 49:769--822, 1916.

\bibitem[Ein17]{Einstein1917}
A.~Einstein.
\newblock Kosmologische Betrachtungen zur Allgemeinen
  Re\-la\-ti\-vi\-t\"{a}ts\-theo\-rie. (German).
\newblock {\em Preu{\ss}. Akad. Wiss., Sitzungsber.}, pp. 142--152, 1917.

\bibitem[Fri22]{Friedmann1922}
A.~Friedmann.
\newblock \"{U}ber die Kr\"{u}mmung des Raumes. (German).
\newblock {\em Z. Phys.}, 10:377--386, 1922.

\bibitem[Fri24]{Friedmann1924}
A.~Friedmann.
\newblock \"{U}ber die M\"{o}glichkeit einer Welt mit konstanter
  negativer Kr\"{u}mmung des Raumes. (German).
\newblock {\em Z. Phys.}, 21:326--332, 1924.

\bibitem[GH96]{GrunewaldHuntebrinker1996}
F.~Grunewald and W.~Huntebrinker.
\newblock A numerical study of eigenvalues of the hyperbolic Laplacian for
  polyhedra with one cusp.
\newblock {\em Experiment. Math.}, 5:57--80, 1996.

\bibitem[GH03]{WMAP2003}
The Wilkinson Microwave Anisotropy Probe Science Team (WMAP) (G.~Hinshaw).
\newblock The NASA Legacy Archive for Microwave Data Analysis
  (LAMBDA), 2003.
\newblock http://lambda.gsfc.nasa.gov/product/map/.

\bibitem[HA93]{HejhalArno1993}
D.~A. Hejhal and S.~Arno.
\newblock On Fourier coefficients of Maass waveforms for
  $\operatorname{PSL}(2,\mathds{Z})$.
\newblock {\em Math. Comp.}, 61:245--267, 1993.

\bibitem[HAB{\etalchar{+}}00]{HananyETal2000}
S.~Hanany, P.~Ade, A.~Balbi, J.~Bock, J.~Borrill, A.~Boscaleri,
  P.~de Bernardis, P.~G. Ferreira, V.~V. Hristov, A.~H. Jaffe, A.~E. Lange,
  A.~T. Lee, P.~D. Mauskopf, C.~B. Netterfield, S.~Oh, E.~Pascale, B.~Rabii,
  P.~L. Richards, G.~F. Smoot, R.~Stompor, C.~D. Winant, and J.~H.~P. Wu.
\newblock MAXIMA-1: A measurement of the cosmic microwave background
  anisotropy on angular scales of 10 arcminutes to 5 degrees.
\newblock {\em Astroph. J.}, 545:L5, 2000.

\bibitem[Hei92]{Heitkamp1992}
D.~Heitkamp.
\newblock Hecke-Theorie zur {${\rm SL}(2;\mathfrak{o})$}. (German).
\newblock {\em Schriftenreihe des Mathematischen Instituts der
  Universit\"{a}t M\"{u}nster, 3. Serie}, 5, 1992.

\bibitem[Hej83]{Hejhal1983}
D.~A. Hejhal.
\newblock {\em The Selberg trace formula for
  $\operatorname{PSL}(2,\mathds{R})$}.
\newblock Lecture Notes in Math. 1001. Springer, 1983.

\bibitem[Hej99]{Hejhal1999}
D.~A. Hejhal.
\newblock On eigenfunctions of the Laplacian for Hecke triangle groups.
\newblock In D.~A. Hejhal, J.~Friedman, M.~C. Gutzwiller, and A.~M. Odlyzko,
  editors, {\em Emerging applications of number theory}, IMA Series No. 109,
  pp. 291--315. Springer, 1999.

\bibitem[Hub29]{Hubble1929}
E.~P. Hubble.
\newblock A relation between distance and radial velocity among extra-galactic
  nebulae.
\newblock {\em Proc. Nat. Acad. Sci. (USA)}, 15:168--173, 1929.

\bibitem[Hum19]{Humbert1919}
G.~Humbert.
\newblock Sur la mesure des classes d'Hermite de discriminant donn\'{e} dans
  un corps quadratique imaginaire, et sur certaines volumes non euclidiens.
  (French).
\newblock {\em C. R. Acad. Sci. Paris}, 169:448--454, 1919.

\bibitem[Hun96]{Huntebrinker1996}
W.~Huntebrinker.
\newblock Numerical computation of eigenvalues of the Laplace-Beltrami
  operator on three-dimensional hyperbolic spaces by finite-element methods.
\newblock {\em Diss. Summ. Math.}, 1:29--36, 1996.

\bibitem[ITS00]{InoueTomitaSugiyama2000}
K.~T. Inoue, K.~Tomita, and N.~Sugiyama.
\newblock Temperature correlations in a compact hyperbolic universe.
\newblock {\em Mon. Not. Roy. Astron. Soc.}, 314:L21, 2000.

\bibitem[KAB{\etalchar{+}}02]{KuoETal2002}
C.~L. Kuo, P.~A.~R. Ade, J.~J. Bock, C.~Cantalupo, M.~D. Daub, J.~Goldstein,
  W.~L. Holzapfel, A.~E. Lange, M.~Lueker, M.~Newcomb, J.~B. Peterson, J.~Ruhl,
  M.~C. Runyan, and E.~Torbet.
\newblock High resolution observations of the CMB power spectrum with
  ACBAR.
\newblock {\em Astroph. J.}, 600:32--51, 2004.

\bibitem[Kub73]{Kubota1973}
T.~Kubota.
\newblock {\em Elementary Theory of Eisenstein Series}.
\newblock Kodansha, Tokyo and Halsted Press, 1973.

\bibitem[LBBS97]{LevinBarrowBunnSilk1997}
J.~J. Levin, J.~D. Barrow, E.~F. Bunn, and J.~Silk.
\newblock Flat spots: Topological signatures of an open universe in cosmic
  background explorer sky maps.
\newblock {\em Phys. Rev. Lett.}, 79:974--977, 1997.

\bibitem[Lem27]{Lemaitre1927}
G.~Lema\^{\i}tre.
\newblock Un univers homog\`{e}ne de masse constante et de rayon croissant,
  rendant compte de la vitesse radiale de n\'{e}buleuses extragalactiques.
  (French).
\newblock {\em Ann. Soc. Sci. Bruxelles}, 47A:47--59, 1927.

\bibitem[Lev02]{Levin2002}
J.~Levin.
\newblock Topology and the cosmic microwave background.
\newblock {\em Phys. Rep.}, 365:251--333, 2002.

\bibitem[Lif46]{Lifshitz1946}
E.~Lifshitz.
\newblock On the gravitational stability of the expanding universe.
\newblock {\em Zh. Eksp. Teor. Fiz.}, 16:587--602, 1946.

\bibitem[Maa49a]{Maass1949a}
H.~Maa{\ss}.
\newblock \"{U}ber eine neue Art von nichtanalytischen automorphen
  Funktionen und die Bestimmung Dirichletscher Reihen durch
  Funk\-tio\-nal\-glei\-chung\-en. (German).
\newblock {\em Math. Ann.}, 121:141--183, 1949.

\bibitem[Maa49b]{Maass1949b}
H.~Maa{\ss}.
\newblock Automorphe Funktionen von mehreren Ver\"{a}nderlichen und
  Dirichletsche Reihen. (German).
\newblock {\em Abh. Math. Semin. Univ. Hamb.}, 16:72--100, 1949.

\bibitem[Mat95]{Matthies1995}
C.~Matthies.
\newblock {\em Picards Billard. Ein Modell f\"{u}r Arithmetisches
  Quantenchaos in drei Dimensionen.} (German).
\newblock PhD thesis, Universit\"{a}t Hamburg, 1995.

\bibitem[Meh91]{Mehta1991}
M.~L. Mehta.
\newblock {\em Random matrices}.
\newblock Academic Press, second edition, 1991.

\bibitem[Mey88a]{Meyerhoff1988a}
R.~Meyerhoff.
\newblock A lower bound for the covolume of hyperbolic 3-orbifolds.
\newblock {\em Duke Math. J.}, 57:185--203, 1988.

\bibitem[Mey88b]{Meyerhoff1988b}
R.~Meyerhoff.
\newblock Sphere packing and volume in hyperbolic 3-space.
\newblock {\em Comment. Math. Helv.}, 61:271--278, 1988.

\bibitem[MFB92]{MukhanovFeldmanBrandenberger1992}
V.~F. Mukhanov, H.~A. Feldman, and R.~H. Brandenberger.
\newblock Theory of cosmological perturbations.
\newblock {\em Phys. Reports}, 215:203--333, 1992.

\bibitem[Mos73]{Mostow1973}
G.~D. Mostow.
\newblock {\em Strong rigidity of locally symmetric spaces}.
\newblock Annals of Mathematics Studies, Princeton University Press, 1973.

\bibitem[Pee65]{Peebles1965}
P.~J.~E. Peebles.
\newblock The black-body radiation content of the universe and the formation
  of galaxies.
\newblock {\em Astroph. J.}, 142:1317--1326, 1965.

\bibitem[PMR{\etalchar{+}}03]{PearsonETal2003}
T.~J. Pearson, B.~S. Mason, A.~C.~S. Readhead, M.~C. Shepherd, J.~L. Sievers,
  P.~S. Udomprasert, J.~K. Cartwright, A.~J. Farmer, S.~Padin, S.~T. Myers,
  J.~R. Bond, C.~R. Contaldi, U.-L. Pen, S.~Prunet, D.~Pogosyan,
  J.~E. Carlstrom, J.~Kovac, E.~M. Leitch, C.~Pryke, N.~W. Halverson,
  W.~L. Holzapfel, P.~Altamirano, L.~Bronfman, S.~Casassus, J.~May, and M.~Joy.
\newblock The anisotropy of the microwave background to l = 3500: Mosaic
  observations with the Cosmic Background Imager (CBI).
\newblock {\em Astroph. J.}, 591:556--574, 2003.

\bibitem[PNB{\etalchar{+}}03]{PageETal2003}
L.~Page, M.~R. Nolta, C.~Barnes, C.~L. Bennett, M.~Halpern, G.~Hinshaw,
  N.~Jarosik, A.~Kogut, M.~Limon, S.~S. Meyer, H.~V. Peiris, D.~N. Spergel,
  G.~S. Tucker, E.~Wollack, and E.~L. Wright.
\newblock First year Wilkinson Microwave Anisotropy Probe (WMAP)
  observations: Interpretation of the TT and TE angular power spectrum
  peaks.
\newblock {\em Astroph. J. Suppl.}, 148:233--241, 2003.

\bibitem[Pra73]{Prasad1973}
G.~Prasad.
\newblock Strong rigidity of $\mathds{Q}$-rank 1 lattices.
\newblock {\em Invent. Math.}, 21:255--286, 1973.

\bibitem[Prz01]{Przeworski2001}
A.~Przeworski.
\newblock Cones embedded in hyperbolic manifolds.
\newblock {\em J. Differential Geom.}, 58:219--232, 2001.

\bibitem[PW65]{PenziasWilson1965}
A.~A. Penzias and R.~W. Wilson.
\newblock A measurement of excess antenna temperature at $4080\,\text{Mc/s}$.
\newblock {\em Astroph. J.}, 142:419--421, 1965.

\bibitem[Roe66]{Roelcke1966}
W.~Roelcke.
\newblock Das Eigenwertproblem der automorphen Formen in der
  hyperbolischen Ebene, I and II. (German).
\newblock {\em Math. Ann.}, 167:292--337, 1966 and 168:261--324, 1967.

\bibitem[Sar95]{Sarnak1995}
P.~Sarnak.
\newblock Arithmetic quantum chaos.
\newblock {\em Israel Math. Conf. Proc.}, 8:183--236, 1995.

\bibitem[SBK{\etalchar{+}}92]{SmootETal1992}
G.~F. Smoot, C.~L. Bennett, A.~Kogut, E.~L. Wright, J.~Aymon, N.~W. Boggess,
  E.~S. Cheng, G.~De Amici, S.~Gulkis, M.~G. Hauser, G.~Hinshaw,
  P.~D. Jackson, M.~Janssen, E.~Kaita, T.~Kelsall, P.~Keegstra, C.~Lineweaver,
  K.~Loewenstein, P.~Lubin, J.~Mather, S.~S. Meyer, S.~H. Moseley, T.~Murdock,
  L.~Rokke, R.~F. Silverberg, L.~Tenorio, R.~Weiss, and D.~T. Wilkinson.
\newblock Structure in the COBE Differential Microwave Radiometer
  first-year maps.
\newblock {\em Astroph. J.}, 396:L1--L5, 1992.

\bibitem[Sel56]{Selberg1956}
A.~Selberg.
\newblock Harmonic analysis and discontinuous groups in weakly symmetric
  Riemannian spaces with applications to Dirichlet series.
\newblock {\em J. Indian Math. Soc.}, 20:47--87, 1956.

\bibitem[SG91]{SmotrovGolovchanskii1991}
M.~N. Smotrov and V.~V. Golov\v{c}anski\v{\i}.
\newblock Small eigenvalues of the Laplacian on $\Gamma\backslash H_3$
  for $\Gamma=PSL_2(\mathds{Z}[i])$.
\newblock {\em Preprint}, 91-040, Bielefeld 1991.

\bibitem[SS76]{SokolovStarobinskii1976}
D.~D. Sokolov and A.~A. Starobinskii.
\newblock Globally inhomogeneous ``spliced'' universes.
\newblock {\em Sov. Astron.}, 19:629--633, 1976.

\bibitem[SS02]{SelanderStrombergsson2002}
B.~Selander and A.~Str\"{o}mbergsson.
\newblock Sextic coverings of genus two which are branched at three points.
\newblock {\em UUDM report 2002:16}, Uppsala 2002.

\bibitem[Sta84]{Stark1984}
H.~M. Stark.
\newblock Fourier coefficients of Maass waveforms.
\newblock In R.~A. Rankin, editor, {\em Modular Forms}, pp. 263--269.
  Ellis Horwood, 1984.

\bibitem[Ste99]{Steil1999}
G.~Steil.
\newblock Eigenvalues of the Laplacian for Bianchi groups.
\newblock In D.~A. Hejhal, J.~Friedman, M.~C. Gutzwiller, and A.~M. Odlyzko,
  editors, {\em Emerging applications of number theory}, IMA Series No. 109,
  pp. 617--641. Springer, 1999.

\bibitem[Str94]{Stramm1994}
K.~Stramm.
\newblock Kleine Eigenwerte des Laplace-Operators zu
  Kongruenzgruppen. (German).
\newblock {\em Schriftenreihe des Mathematischen Instituts der
  Universit\"{a}t M\"{u}nster, 3. Serie}, 11, 1994.

\bibitem[SVP{\etalchar{+}}03]{SpergelETal2003}
D.~N. Spergel, L.~Verde, H.~V. Peiris, E.~Komatsu, M.~R. Nolta, C.~L. Bennett,
  M.~Halpern, G.~Hinshaw, N.~Jarosik, A.~Kogut, M.~Limon, S.~S. Meyer,
  L.~Page, G.~S. Tucker, J.~L. Weiland, E.~Wollack, and E.~L. Wright.
\newblock First year Wilkinson microwave anisotropy probe (WMAP)
  observations: Determination of cosmological parameters.
\newblock {\em Astroph. J. Suppl.}, 148:175--194, 2003.

\bibitem[SW67]{SachsWolfe1967}
R.~K. Sachs and A.~M. Wolfe.
\newblock Perturbations of a cosmological model and angular variations of the
  microwave background.
\newblock {\em Astroph. J.}, 147:73--90, 1967.

\bibitem[The02]{Then2002}
H.~Then.
\newblock Maass cusp forms for large eigenvalues.
\newblock Accepted for publication in Math. Comp.,
\newblock {\em math-ph/0305047}.

\bibitem[The03]{Then2003}
H.~Then.
\newblock Arithmetic quantum chaos of Maass waveforms.
\newblock In B.~Julia, P.~Moussa, P.~Cartier, and P.~Vanhove,
  editors, {\em Number Theory, Physics, and Geometry}. To be published
  by Springer.
\newblock {\em math-ph/0305048}.

\bibitem[Wat44]{Watson1944}
G.~N. Watson.
\newblock {\em A treatise on the theory of Bessel functions}.
\newblock Cambridge University Press, 1944.

\bibitem[Wee85]{Weeks1985}
J.~Weeks.
\newblock {\em Hyperbolic structures on 3-manifolds}.
\newblock PhD thesis, Princeton University, 1985.

\bibitem[Wey12]{Weyl1912}
H.~Weyl.
\newblock Das asymptotische Verteilungsgesetz der Eigenwerte linearer
  partieller Differentialgleichungen. (German).
\newblock {\em Math. Ann.}, 71:441--479, 1912.

\end{thebibliography}
\end{document}